\documentclass[%
superscriptaddress,
 amsmath,amssymb,
 aps,
 jcp,
 letterpaper,
]{revtex4-2}

\usepackage{graphicx}
\usepackage{dcolumn}
\usepackage{bm}

\newcolumntype{.}{D{.}{.}{8}}

\usepackage[utf8]{inputenc}
\usepackage{physics}
\usepackage{amsmath, amssymb}
\usepackage{tabularx,booktabs,array,dcolumn}
\usepackage{setspace}
\usepackage{vmargin}
\usepackage{xcolor}
\usepackage{graphicx,psfrag,subfigure}

\usepackage{float}
\usepackage[all]{xy}
\usepackage{multirow}

\usepackage{bm}
\usepackage{mathtools}
\usepackage{physics}

\newcommand{\bos}[1]{\boldsymbol{#1}}

\newcommand{\cm}{cm$^{-1}$}

\newcommand{\som}{SOM}

\usepackage[unicode]{hyperref}
\hypersetup{
   unicode=true,          
   plainpages=false,
   colorlinks=true,       
   citecolor=blue,        
}

\def\dd{\text{d}}
\def\tor{\tau}   
\def\cc{c-}   
\def\rc{rc-}  
\def\ac{ac-}  
\def\HO{\text{HO}}
\def\tmpes{TM16-PES}
\def\eq{\text{eq}}
\def\mL{\mathcal{L}}
\def\mQ{\mathcal{Q}}
\def\deg{\text{o}}
\def\tD{{\bar{D}}}
\def\tmL{\tilde{\mathcal{L}}}
\def\tmQ{\tilde{\mathcal{Q}}}
\def\nmax{N^\text{max}}
\def\mF{\mathcal{F}}
\def\gesm{GENIUSH-Smolyak}
\def\trans{\emph{trans}}
\def\cis{\emph{cis}}
\def\delocalized{\emph{delocalized}}

\usepackage{soul}
\usepackage{color}
\definecolor{fgreen}{rgb}{0.0, 0.5, 0.0}

\begin{document}

\title{%
Variational vibrational states of HCOOH
}

\author{Alberto Mart\'in Santa Dar\'ia}

\author{Gustavo Avila}
\email{Gustavo$_$Avila@telefonica.net}

\author{Edit M\'atyus}
\email{edit.matyus@ttk.elte.hu}

\affiliation{
ELTE, E\"otv\"os Lor\'and University,
Institute of Chemistry, 
P\'azm\'any P\'eter s\'et\'any 1/A, 
1117 Budapest, Hungary}

\date{11 March 2022}
\begin{abstract}
  \noindent 
  Vibrational states of the formic acid molecule are converged using the GENIUSH--Smolyak approach 
  and the potential energy surface taken from [D. Tew and W. Mizukami, J.~Phys.~Chem.~A 120, 9815 (2016)].
  The quantum nuclear motion is described by using the \cis-\trans\ torsional coordinate and eight curvilinear normal coordinates defined with respect to an instantaneous reference configuration changing as a function of the torsional degree of freedom.
  Harmonic oscillator basis functions are used for the curvilinear normal coordinates, a Fourier basis for the torsional coordinate, and a simple basis pruning condition is combined with a Smolyak integration grid.
  \emph{Trans,} \cis, and \emph{delocalized} vibrational states are reported up to and slightly beyond the isomerization barrier.

  \begin{center}
    \includegraphics[scale=0.5]{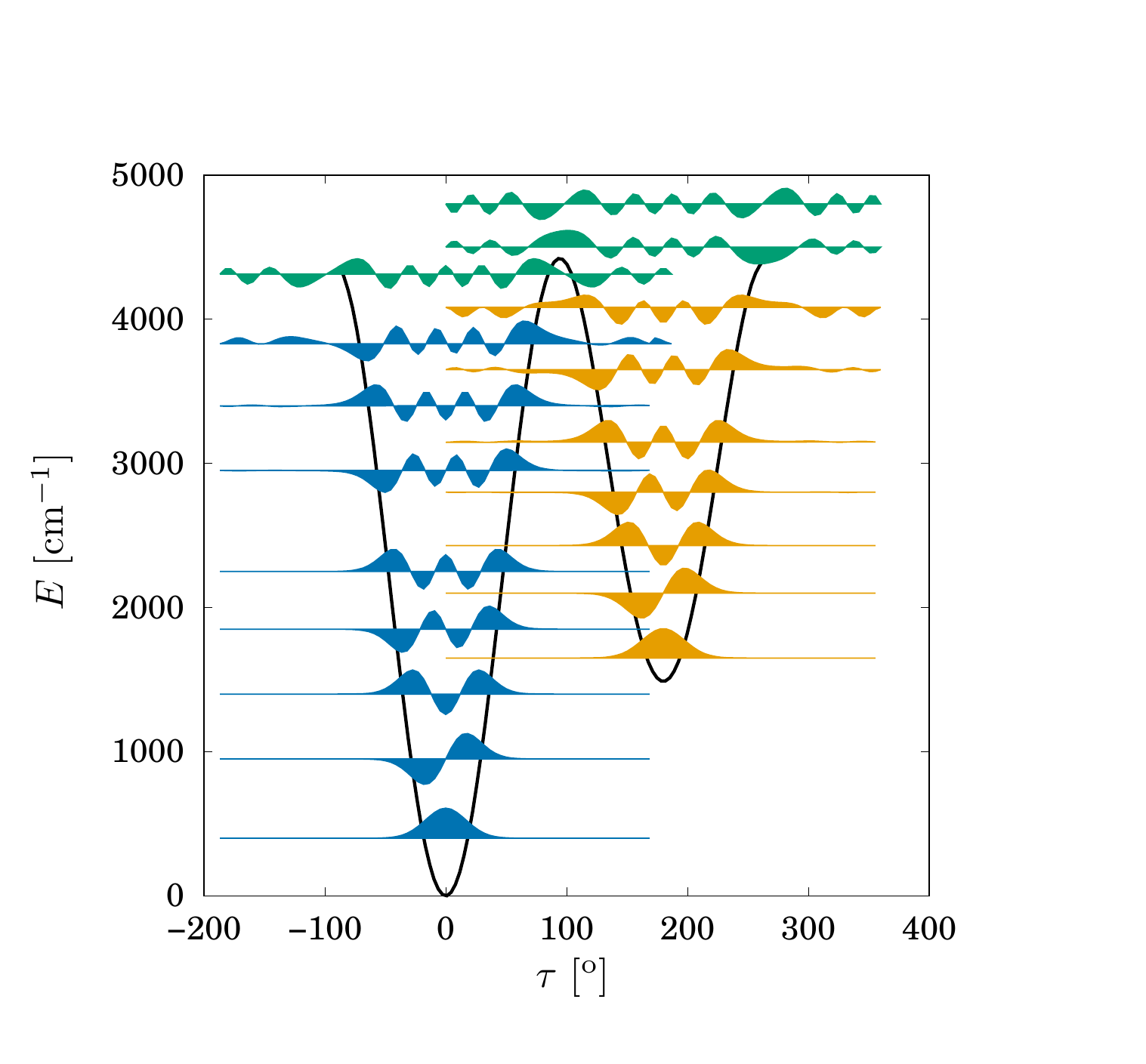}
  \end{center}
\end{abstract}

\maketitle

%
%
\clearpage
\section{Introduction}
\noindent 
This paper is dedicated to the memory of J. K. G. Watson, the father of the Watson Hamiltonian,
a compact, analytic formulation of the rovibrational kinetic energy operator in normal coordinates. 
In this paper, we also use normal coordinates, but we adapt these coordinates to the \cis-\trans\ isomerization of the formic acid molecule. There is no known analytic formulation for the corresponding kinetic energy operator available, and we use computer power to construct the kinetic energy coefficients where they are needed. 

Over the past decade, we have observed a rapid and fundamentally important development of exact quantum dynamics techniques to solve the (ro)vibrational problem.
Development is observed in several directions:
(a)~coordinate representation and the kinetic energy operator \cite{LaNa02,MaCzCs09,FaMaCs11,14FaMaCs,YaYu15,LaNa14,NaLa18};
(b)~contraction techniques \cite{WaCa18,WaCa20vinyl,FeBa19,FeBa20,LiLiFeBa21};
(c)~grid pruning techniques \cite{AvCa09,AvCa11,AvCa11b,AvMa19,AvMa19b,ChLa21};
(d)~collocation \cite{WoCa19,WoCa21,Ca21};
(e)~accurate potential energy representations for high-dimensional systems \cite{PeSaMe14,OtChiPe18,RaPe20}; 
(f)~highly parallel computation of ten thousands or millions of vibrational states \cite{HaPo15,HaPo15b,SaPo21}.

Regarding the formic acid molecule, 
there are two full-dimensional, high-level \emph{ab initio} potential energy surfaces (PESs) \cite{TeMi16,RiCa18} that have been used in sophisticated (variational or perturbative) vibrational computations. 
Tew and Mizukami used their PES in a variational vibrational computation with a five-mode representation and the  internal-coordinate path Hamiltonian (ICPH) approach \cite{TeMi16}.
Richter and Carbonni\`ere  used a similar PES \cite{RiCa18}, computed vibrational energies using a valence-coordinate representation of the kinetic energy operator and the multi-configuration time-dependent Hartree approach, and they reported significant deviations from Ref.~\cite{TeMi16} for the vibrational states of the \cis\ potential energy well. 
Last year, Nejad and Sibert used both PESs and sixth-order canonical Van Vleck perturbation theory (CVPT) in curvilinear normal coordinates localized in one of the potential energy wells (\trans\ or \cis) of the molecule \cite{NeSi21}.

In the present work, we focus on the vibrational methodology and define an efficient setup that can be used to converge (better than 5~\cm) all vibrational states of the formic acid molecule up to and possibly beyond the isomerization barrier. 
During the course of the development of a benchmark-quality variational vibrational setup, we use the Tew--Mizukami PES \cite{TeMi16} (henceforth labelled as \tmpes). It is left for future work, when well-converged vibrational energies can be `routinely' computed for the relevant energy range of this system, to decide which PES representation performs better in comparison with experiment (gas-phase overtone and combination bands). 
Vibrational band origins are available from experimental infrared and Raman observations, a review and an extensive list of references can be found in the introduction of Ref.~\cite{NeSi21}.

The present work is organized as follows. 
Sec.~\ref{ch:coor} reports the development of a torsional-curvilinear normal coordinate representation.  
Sec.~\ref{ch:keo} describes the construction of the corresponding kinetic energy operator coefficients using the numerical kinetic energy operator approach as it is implemented in the GENIUSH computer program \cite{MaCzCs09}.
Sec.~\ref{ch:basis} defines the harmonic oscillator basis functions for the curvilinear normal coordinates, the Fourier basis and torsional functions for the torsional degree of freedom, and discusses basis pruning strategies.
Sec.~\ref{ch:multigrid} describes the Smolyak non-product grid technique that is used to compute multi-dimensional integrals.
In Sec.~\ref{ch:numres}, vibrational energies are presented and discussed in relation with earlier computations \cite{TeMi16,NeSi21}, and further necessary development and computational work is outlined.

\begin{figure}
    \centering
    \includegraphics[scale=0.6]{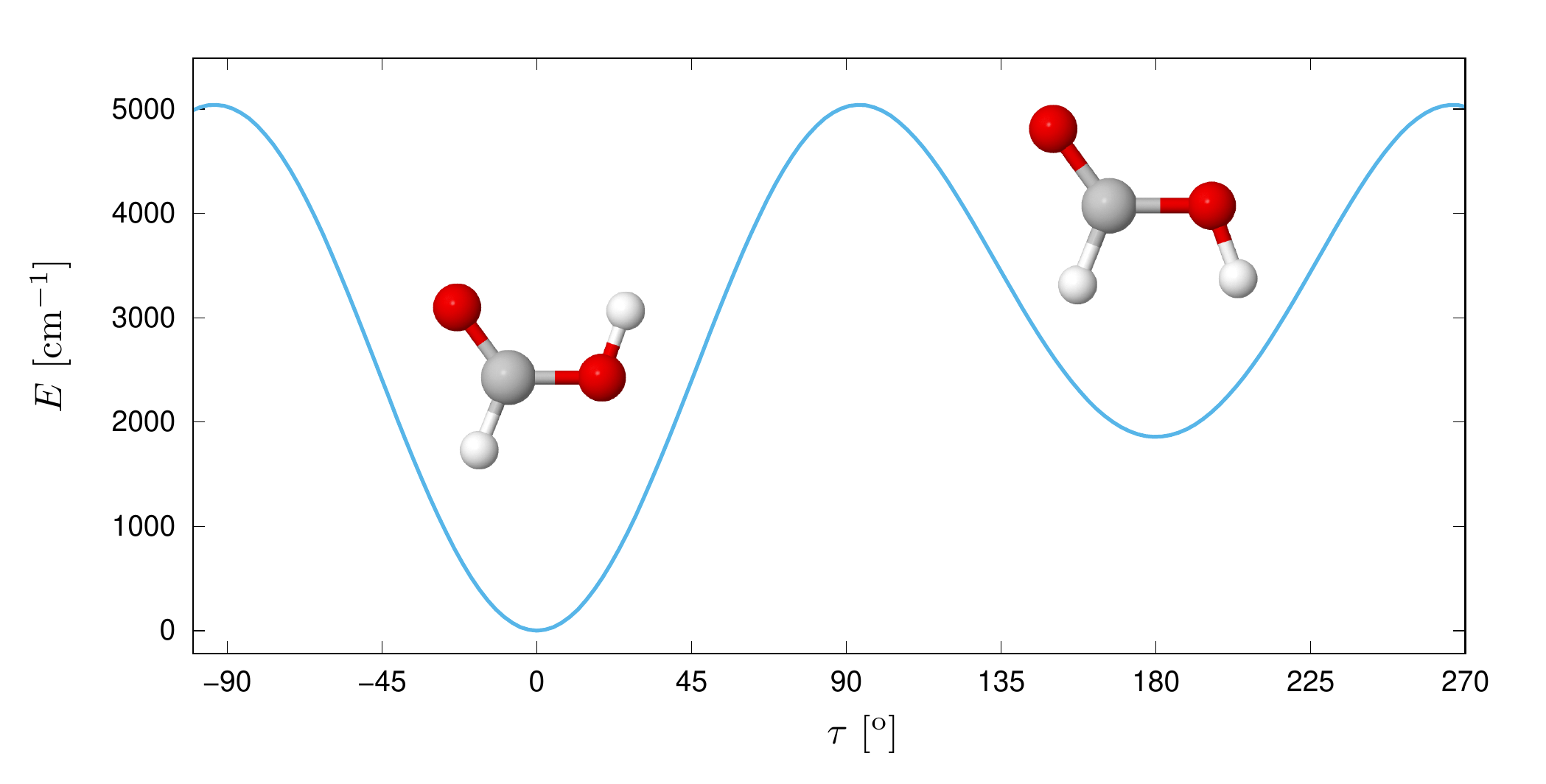}
    \caption{1-dimensional cut of the \tmpes\ \cite{TeMi16} along the torsional angle ($\tor$) that describes the conversion between the \trans\ and the \cis\ conformers of the formic acid molecule.}
    \label{fig:pector}
\end{figure}

\section{Vibrational coordinates \label{ch:coor}}
\subsection{Internal coordinates}
\begin{figure}
  \begin{center}
    \includegraphics[scale=0.25]{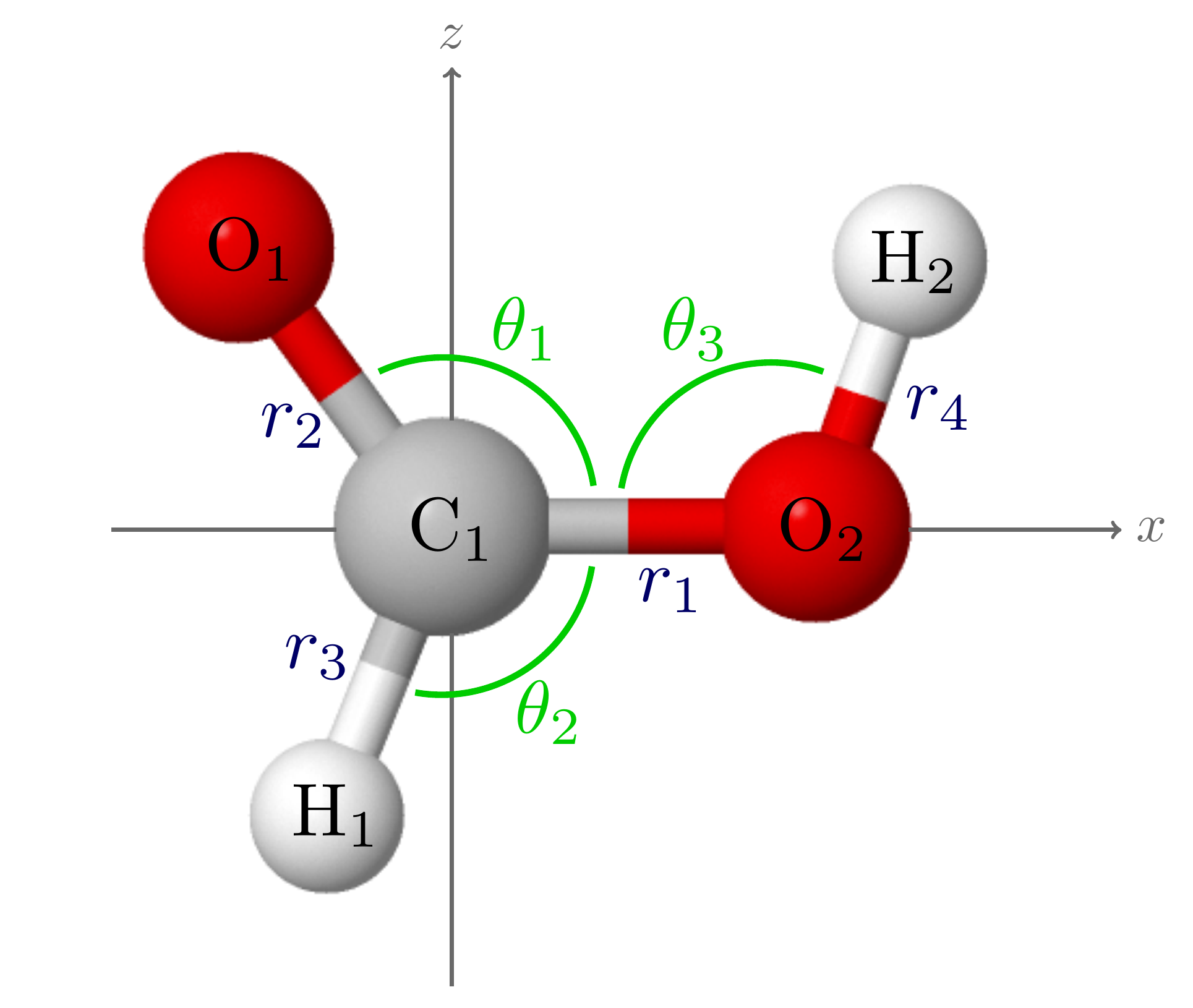}
    \vspace{0.5cm}
    \begin{tabular}{@{} ll@{\ \ } l@{\ \ } l@{\ \ } l@{\ \ } l@{\ \ } l@{\ \ } l@{\ \ }@{}}
      \hline\\[-0.40cm]
      \hline\\[-0.40cm]
      1 & C$_1$ & &   &   &   &   &   \\ 
      2 & O$_2$ & C$_1$   & $r_1$ [1.342]$^\text{t}$[1.349]$^\text{c}$ &   &   &   &   \\ 
      3 & O$_1$ & C$_1$   & $r_2$ [1.199]$^\text{t}$[1.192]$^\text{c}$ & O$_2$  & $\theta_1$ [124.9]$^\text{t}$[122.3]$^\text{c}$ &   &   \\ 
      4 & H$_1$ & C$_1$   & $r_3$ [1.094]$^\text{t}$[1.100]$^\text{c}$ & O$_2$  & $\theta_2$ [110.0]$^\text{t}$[113.7]$^\text{c}$ & O$_1$  & $\varphi$ [0.0]$^\text{t}$[0.0]$^\text{c}$ \\ 
      5 & H$_2$ & O$_2$   & $r_4$ [0.967]$^\text{t}$[0.962]$^\text{c}$ & C$_1$  & $\theta_3$ [106.7]$^\text{t}$[109.2]$^\text{c}$ & O$_1$  & $\tor$ [0.0]$^\text{t}$[180.0]$^\text{c}$ \\ [0.cm]
      \hline\\[-0.40cm]
      \hline
    \end{tabular}
    \vspace{-1.cm}
    \end{center}      
    \caption{%
      Visualization of the internal coordinates, Eq.~(\ref{eq:internal}), for the example of the \trans-formic acid molecule in its equilibrium structure. 
      The equilibrium values of the distances, in $\textup{\r{A}}$ and the angles, in degree, corresponding to the \cis\ ($^\text{c}$) and the \trans\ ($^\text{t}$) conformers on the \tmpes\ are shown in brackets.
      }
    \label{fig:coor}
\end{figure}

The body-fixed Cartesian coordinates of HCOOH are defined in terms of the $r_i\in[0,\infty)$ distances, the $\theta_i\in[0,\pi]$ angles, the $\varphi\in[-\pi,\pi)$ out-of-plane bending, and the $\tau\in[0,2\pi)$ torsional angle according to the following expressions:
\begin{equation}
\begin{split}
    {\bos{r}}_{\text{C}_1}= \bos{0} \; ,
    \quad\quad \bos{r}_{\text{O}_2}= 
\begin{pmatrix}
0\\ 
0\\
r_{1}
\end{pmatrix} \; ,
\quad\quad \bos{r}_{\text{O}_1}= 
\begin{pmatrix}
0\\
r_{2} \cos{(\theta_{1}-\pi/2)} \\
-r_{2} \sin{(\theta_{1}-\pi/2)}
\end{pmatrix} \; ,
\\
\\
\bos{r}_{\text{H}_1}= 
\begin{pmatrix}
r_{3} \cos{(\theta_{2}-\pi/2)} \sin\varphi\\
-r_{3} \cos{(\theta_{2}-\pi/2)} \cos\varphi\\
-r_{3} \sin{(\theta_{2}-\pi/2)}
\end{pmatrix} \; ,
\quad\quad \bos{r}_{\text{H}_2}= \bos{r}_{\text{O}_2} + 
\begin{pmatrix}
r_{4} \cos{(\theta_{3}-\pi/2)} \sin\tau \\
r_{4} \cos{(\theta_{3}-\pi/2)} \cos\tau \\
r_{4} \sin{(\theta_{3}-\pi/2)}
\end{pmatrix} \; .
\end{split}
\label{eq:internal}
\end{equation}
The coordinate definition and the corresponding $Z$-matrix including the equilibrium values at the \trans\ and the \cis\ minima of the \tmpes\ are summarized in Figure~\ref{fig:coor}. For later use, a compact notation of the coordinates is introduced as
\begin{align}
  \bos{\xi}=(\xi_1,\ldots,\xi_9)=(r_1,r_2,r_3,r_4,\theta_1,\theta_2,\theta_3,\varphi,\tau) \; .
\end{align}

\subsection{Rectilinear normal coordinates}
The body-fixed Cartesian coordinates and displacements with respect to the $c_{i\alpha}$ reference (equilibrium) geometry of an $N$-atomic molecule can be written in terms of the $Q_k\in\mathbb{R}$ normal coordinates as
\begin{align}
  r_{i\alpha} &= c_{i\alpha} + d_{i\alpha} \quad\text{with} \nonumber \\
  d_{i\alpha} &= \frac{1}{m^{1/2}_i}\sum^\tD_{k=1} l_{i\alpha,k}Q_k \; , \quad i = 1,\ldots,N, \ \text{and}\ \alpha=x,y,z \; . 
    \label{eq:normalcoor1}  
\end{align}
The $l_{i\alpha,k}$ coefficients are the elements of the eigenvectors of the $\bos{GF}$ matrix \cite{WiDeCr80} evaluated at the reference (equilibrium) structure. 
In this work, the normal coordinate calculation has been performed at both minima (\trans\ and \cis) of HCOOH, hence, there are two parameter sets. 
The $\bos{F}$ Hessian matrix has been computed by finite differences of the PES at both (\trans\ and \cis) equilibrium structures with respect to the displacements along the $3N$ Cartesian coordinates. 
The numerical derivatives and related mathematical manipulations were evaluated using the Wolfram Mathematica symbolic algebra program \cite{WolframMath}.

\begin{figure}
    \centering
    \includegraphics[scale=0.45]{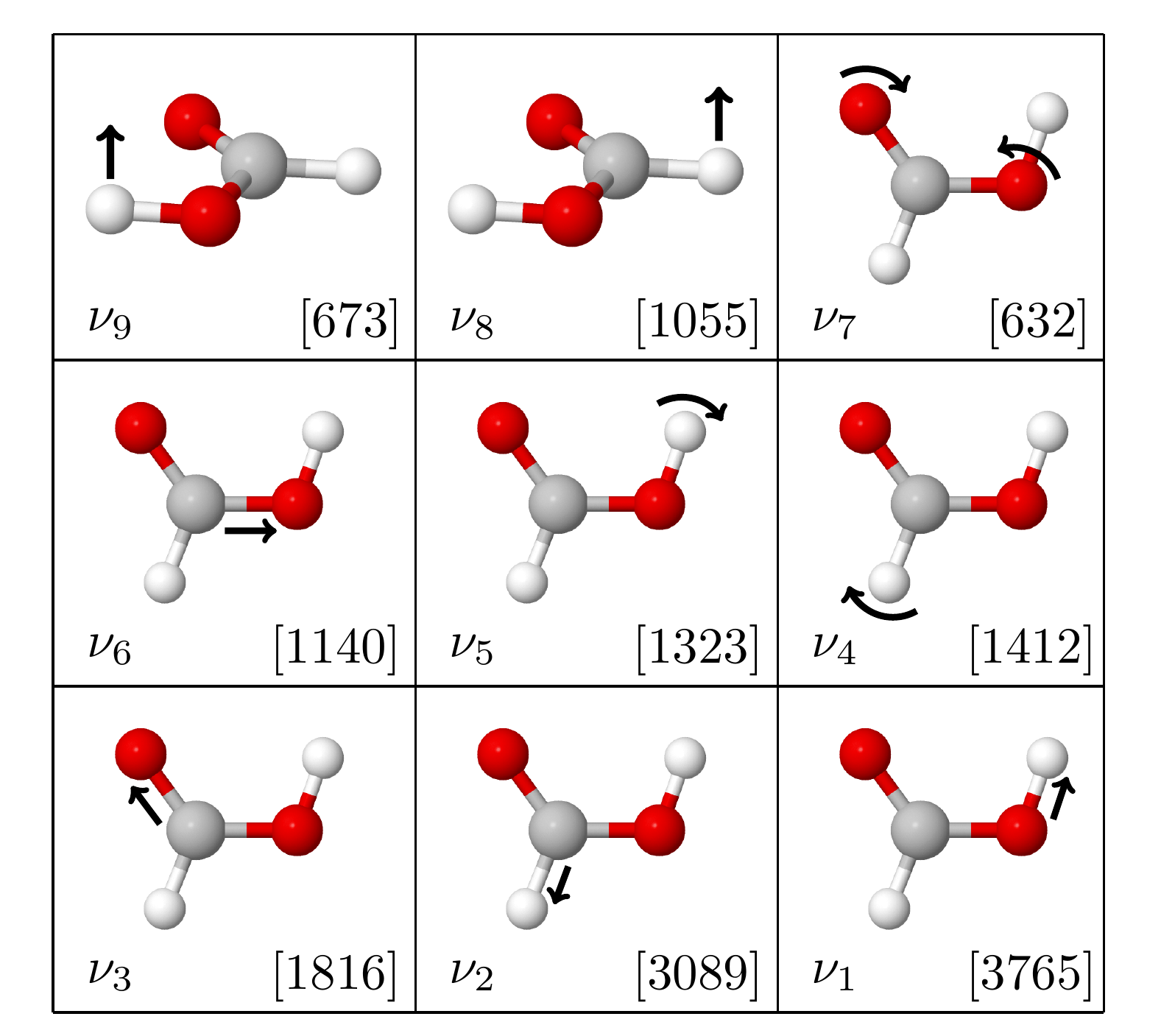}
    \caption{%
      Visualization of the normal modes corresponding to the global minimum (\trans) of HCOOH.  
      The harmonic frequencies, in \cm, are shown in brackets.
    \label{fig:nctrans}
    }
\end{figure}

HCOOH is an $N=5$-atomic molecule and its total number of vibrational degrees of freedom is $3N-6=9$. We used Eq.~(\ref{eq:normalcoor1}) with $\tD=9$ (Fig.~\ref{fig:nctrans}), but we also used it with $\tD=8$ while the $\tor$ torsional degree of freedom was excluded from the harmonic analysis that is necessary to have a good description for the \cis-\trans\ isomerization.

\subsection{Curvilinear normal coordinates \label{ch:curvi}}
Instead of using rectilinear displacement coordinates, $d_{i\alpha}$, 
a better vibrational representation can be obtained \cite{PaAlBook82}, if we use (curvilinear) internal coordinates (Fig.~\ref{fig:coor}), for which curvilinear displacement (\cc-displacement) coordinates can be defined with respect to some reference (equilibrium, eq) value as
\begin{align}
  \Delta \xi_i = \xi_i - \xi_i^{(\eq)} \; .
\end{align}
We define linear combinations of these curvilinear displacement coordinates, and we call them curvilinear (\cc-normal) normal coordinates, 
\begin{align}
  \mathcal{Q} = \bos{\mL}^{-1} (\bos{\Delta\xi})
\end{align}
such that the kinetic and potential energy coupling (near the reference structure) is reduced.
Hence, similarly to the rectilinear normal coordinates, the linear combination coefficients can be obtained using the $\bos{GF}$ method:
\begin{equation}
    \bos{\mL}^{-1} \bos{GF} \bos{\mL} = \bos{\Lambda} 
\end{equation}
and the eigenvectors, in $\bos{\mL}$, of the $\bos{GF}$ matrix provide us the linear combination coefficients of the \cc-normal coordinates.
$\bos{F}$ is the Hessian matrix of the PES
\begin{equation}
    F_{ij} = \bigg(\frac{\partial^2 V}{\partial \xi_i \partial \xi_j}\bigg)_\text{eq}
\end{equation}
computed with respect to the $\xi_i$ curvilinear coordinates at the equilibrium (eq) structure.
The matrix $\bos{G}$ is obtained as 
\begin{align}
    \bos{G} = \bos{B} \bos{M}^{-1} \bos{B}^{T} \; ,
\end{align}
where $\bos{M}$ is a $3N \times 3N$ diagonal matrix containing the masses, $m_a\ (a=1,..., N)$, of the atomic nuclei 
\begin{equation}
    \bos{M}^{-1} = 
\begin{pmatrix}
1/m_1 &  &  &  & 0\\
 & 1/m_1 &  &  & \\
  & & 1/m_1 &  & \\
  & &  & \ddots & \\
0 & &  &  & 1/m_N
\end{pmatrix}
\end{equation}
and $\bos{B}$ is a $\tD\times3N$ matrix that contains the derivatives of the internal coordinates with respect to the rectilinear displacements $(\partial \xi_i/ \partial d_{j\alpha})_{\eq}$ 
($i=1,\ldots,\tD$, $j=1,\ldots,N$, $\alpha=x,y,z$)
and satisfy the following relation:
\begin{equation}
\begin{pmatrix}
\xi_{1}\\
\xi_{2}\\
\vdots\\
\xi_{\tD}
\end{pmatrix}
= \bos{B} 
\begin{pmatrix}
d_{1x}\\
d_{1y}\\
d_{1z}\\
\vdots\\
d_{Nz} 
\end{pmatrix} \; .
\end{equation}

To have a good description of the \cis-\trans\ torsional motion, we exclude the $\tor$ torsional degree of freedom from the $\bos{GF}$ calculation and the $c$-normal coordinates are defined for the remaining $\tD=8$ (displacement) internal coordinates that exhibit small(er) amplitude motions. 

For a given value of the $\mQ_k$ \cc-normal coordinates, 
the value of the $\xi_i$ internal coordinates can be calculated using the eigenvectors in $\bos{\mL}$
and the $\xi_i^{(\eq)}$ equilibrium values:  
\begin{align}
  \xi_i
  =
  \xi_i^{(\eq)}
  + 
  \sum_{k=1}^{\tD}
  \mL^{(\eq)}_{i,k} \mQ_k \; ,\quad k=1,\ldots,\tD(=8)
\end{align}

Since HCOOH has two equilibrium configurations, we have two parameter sets:
$\lbrace \bos{\xi}^\text{(c)},\mL^\text{(c)}\rbrace$ 
and 
$\lbrace \bos{\xi}^\text{(t)}, \mL^\text{(t)}\rbrace$ 
(the parameters are deposited in the \som).

\subsection{Relaxed curvilinear normal coordinates along the torsional motion}
\begin{figure}
    \centering
    \includegraphics[scale=0.6]{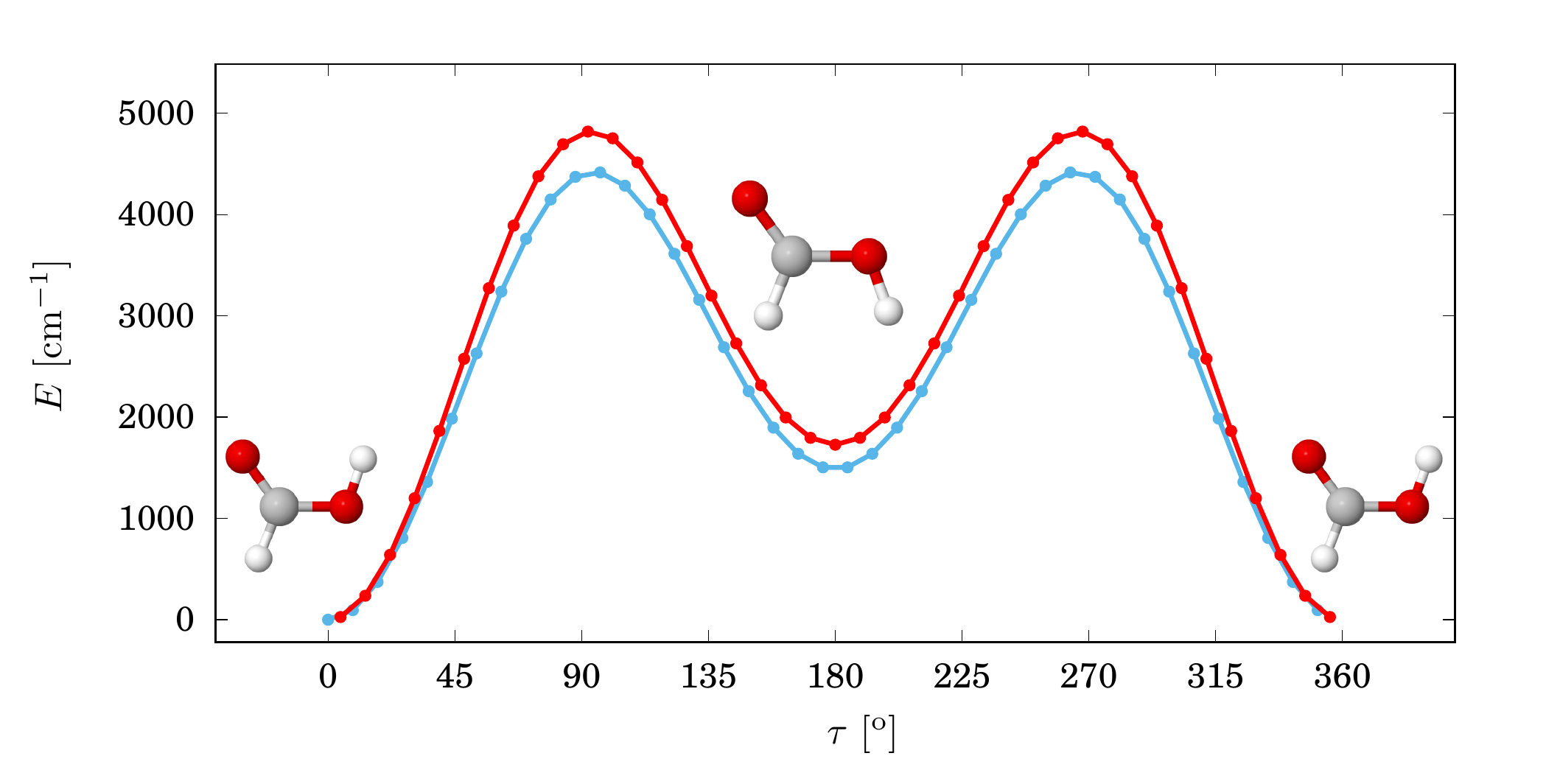} \\[-0.5cm]
    \caption{%
      1-dimensional cut of the PES along the $\tor$ torsional coordinate with (a) the non-torsional coordinates fixed at their equilibrium value at the global minimum (red); (b) relaxed non-torsional coordinates minimizing the potential energy (blue).
    }
    \label{fig:optpes}
\end{figure}

\begin{figure}
    \centering
    \includegraphics[scale=0.33]{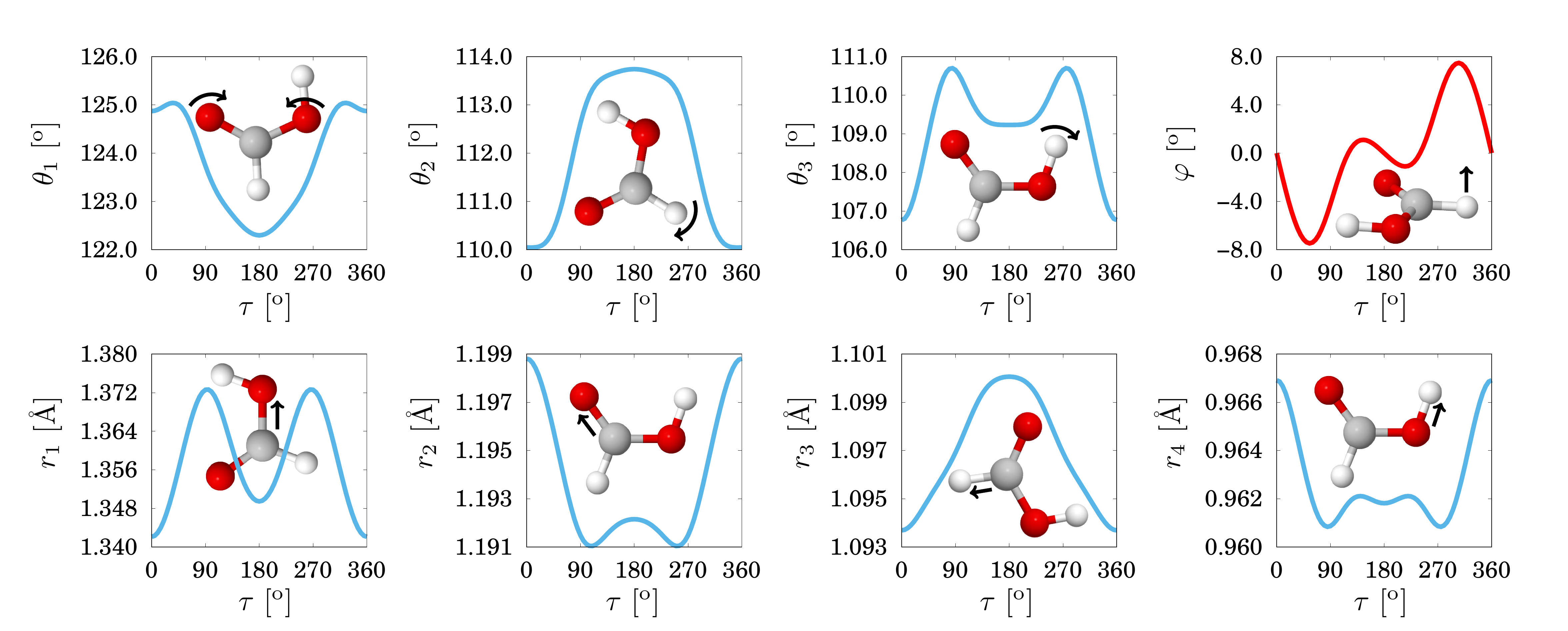}
    \caption{%
    Minimized equilibrium values for the internal coordinates as a function of the $\tor$ torsional angle. The relaxed bond length and bond angle functions are symmetric with respect to $\tor=180^{\circ}$ (blue).
    The relaxed out-of-plane bending angle, $\varphi$, is anti-symmetric with respect to $\tor=180^{\circ}$ (red).
    }
    \label{fig:intcoord}
\end{figure}

Since we are interested in the overall vibrational dynamics of HCOOH, we cannot restrict the description to the \cis\ or the \trans\ well. 
Hence, we repeated the \cc-normal mode computation (with $\tD=8$ degrees of freedom) along the large-amplitude motion at several points \emph{(vide infra)} over the entire range of $\tau\in[0,2\pi)$.
During this computation, we relaxed the molecular structure along $\tau$ by minimizing the potential energy (Fig.~\ref{fig:optpes}). The relaxed values of the internal coordinates that correspond to the minimal potential energy as a function of $\tau$ are shown in Fig.~\ref{fig:intcoord}. These relaxed internal coordinate structures are considered as the `equilibrium structure', $\bos{\xi}^{(\text{eq})}(\tau)$, for the 8-dimensional \cc-normal-mode computation that is repeated for several $\tau$ values.

In practice, the \cc-normal coordinates are computed (similarly to Sec.~\ref{ch:curvi}) 
at 24 equally distributed values of $\tau=t_n$ with $t_n=(n-1) 360^\deg/24$, 
\begin{align}
  \xi_i(\tau)
  = 
  \xi^{\text{(eq)}}_i(\tau) 
  + 
  \sum_{j=1}^8
    \tmL_{i,j} (\tau)
    \tmQ_j  , 
    \quad i=1,\ldots,8 \; .
  \label{eq:rcnormal}
\end{align}
Using the value of the coefficients at the $t_n$ ($n=1,\ldots,24$) points, we interpolate $\tmL_{i,j}(\tor)$ by solving a system of linear equations,
\begin{align}
  \tmL_{i,j}(t_n) 
  = 
  \sum_{k=1}^{24} 
  c_k^{ij} f_k(t_n) \; , \quad n=1,\ldots,24 \; .
\end{align}
For the $f_k$ functions, we have considered the following functions of the Fourier basis
\begin{align}
  1, \cos(\tor), \sin(\tor), \cos(2\tor), \sin(2\tor), \ldots, \cos(12\tau) \; . 
  \label{eq:phitau}
\end{align}
All  coordinates can be expanded using only cosine functions of $\tau$, except for the $\varphi$ out-of-plane bending (that is also a torsion-like, but small-amplitude vibration). $\varphi$ is an odd function of $\tau$ (Fig.~\ref{fig:intcoord}), and hence sine basis functions are used for its interpolation.

\begin{figure}
    \centering
    \includegraphics[scale=1.3]{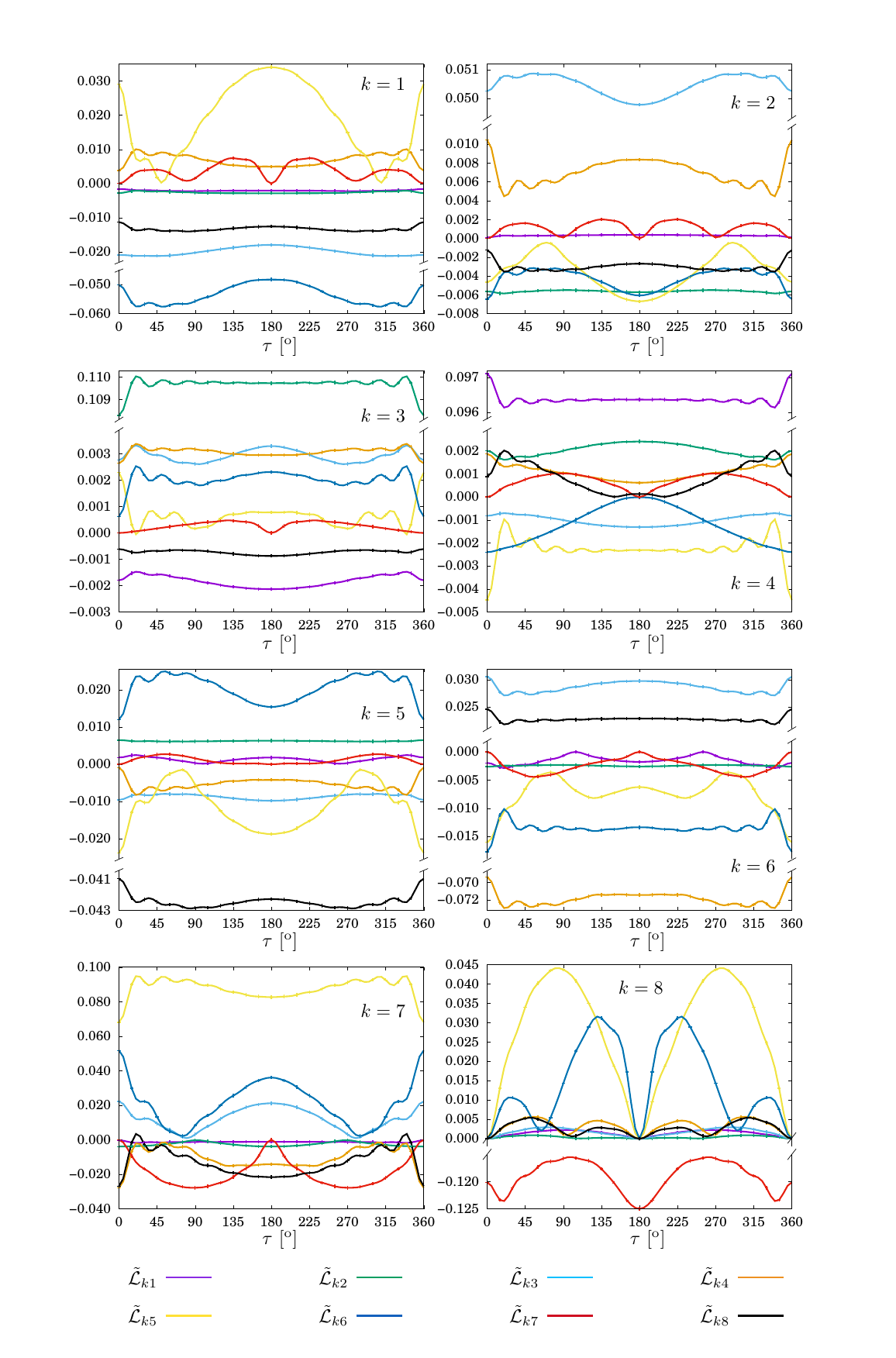}\\[-0.5cm]
    \caption{%
      Curvilinear normal coordinate coefficients, Eq.~(\ref{eq:rcnormal}),
      as a function of the $\tor$ torsional angle. 
    }
    \label{fig:Lcoef}
\end{figure}

As a result, we have relaxed equilibrium internal coordinates, $\bos{\xi}^{(\eq)}(\tau)$ and 
relaxed c-normal (\rc-normal) mode $\tmL_{i,j}(\tau)$ coefficients as a function of $\tau$ (Figs.~\ref{fig:intcoord} and \ref{fig:Lcoef}).
The \rc-normal coordinates incorporate in the coordinate definition the optimal structural changes along the $\tau$ large-amplitude motion, while the kinetic and potential energy coupling is minimized among the small amplitude (normal) coordinates. This construct is expected to provide an almost ideal coordinate representation for this system. Results of convergence tests are reported in Sec.~\ref{ch:numres}.

\section{Quantum Hamiltonian \label{ch:keo}}
\begin{figure}
    \centering
    \includegraphics[scale=0.6]{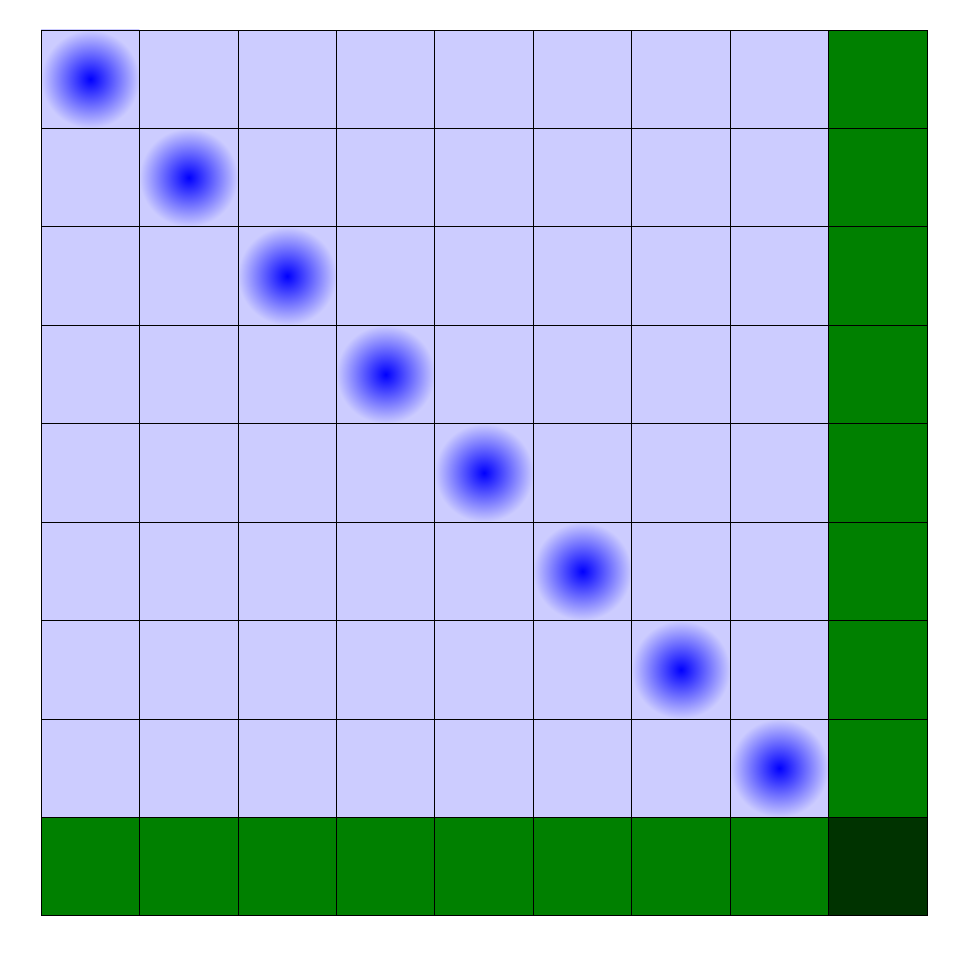}
    \caption{%
      Visualization of the $\bos{G}$ matrix, Eqs.~(\ref{eq:gmxt}) and (\ref{eq:bigGmat}), for a (any) grid point from the dynamically relevant range. The eight \rc-normal modes are weakly coupled among each other (light blue), but their weak coupling 
      with the large amplitude motion ($\tor$) is \emph{not} assumed (dark green). 
      The relaxation of the reference structure and the $\bos{GF}$ diagonalization along $\tor$ ensures that the coupling among the \rc-normal coordinates remains small (light blue) for any value of $\tor$ (see also Figs.~\ref{fig:intcoord} and \ref{fig:Lcoef}).
    }
    \label{fig:keocpl}
\end{figure}

\noindent%
The (ro)vibrational kinetic energy operator (KEO) corresponding to the torsional-relaxed-curvilinear-normal coordinate  representation $\bos{q}=(\bos{\tmQ},\tau)$ is constructed using the numerical KEO approach as implemented in the GENIUSH computer program \cite{MaCzCs09}.
The core of the program is based on the evaluation of the mass-weighted metric tensor at coordinate points:
\begin{equation}
  g_{kl}= \sum_{i=1}^N m_i \textbf{t}^\text{T}_{ik} \textbf{t}_{il}; 
   \quad\quad\quad k,l = 1,2,...,3N-3 
  \label{eq:gmxt}
\end{equation}
where the so-called vibrational and rotational t-vectors are 
\begin{equation}
\textbf{t}_{ik} = \frac{\partial \textbf{r}_i}{\partial q_k}; 
   \quad\quad\quad k = 1,2,...,3N-6
   \label{eq:tvib}
\end{equation}
\begin{equation}
\textbf{t}_{i,3N-6+a} = \textbf{e}_a \times \textbf{r}_i; 
   \quad\quad\quad a = 1(x),2(y),3(z) \; ,
   \label{eq:trot}   
\end{equation}
respectively. 
For the computation of the t-vectors, and using them to construct the $\bos{g}\in\mathbb{R}^{(3N-3)\times(3N-3)}$ matrix, it is necessary to know the body-fixed Cartesian coordinates $\bos{r}_i\ (i=1,\ldots,N)$ as a function of the generalized vibrational coordinates $q_k\ (k=1,\ldots,3N-6)$. 
We expect that an efficient representation can be obtained with the relaxed-curvilinear-normal coordinate plus torsion choice (Sec.~\ref{ch:coor})
\begin{align}
  q_i&=\tmQ_i\; ,\quad i=1,\ldots,8 \nonumber \\
  q_{3N-6}&=\tau \;  .
\end{align}
The corresponding $\bos{r}_i$ vs. $q_k$ relations can be obtained from Eqs.~(\ref{eq:internal}) and (\ref{eq:rcnormal}).
This coordinate choice results in an arrow-like structure of the $\bos{G}$ matrix (Fig.~\ref{fig:keocpl}), \emph{i.e.,}
the coupling of the $\tau$ large-amplitude motion and the $\bos{\tmQ}$ small-amplitude coordinates is not necessarily small (can be large), but the coupling among the small-amplitude $\bos{\tmQ}$ coordinates is small for all $\tau$ values.

The derivatives of the $\bos{r}_i$ Cartesian coordinates with respect the $q_k$ generalized internal coordinates are obtained by using the two-sided finite difference formula. The $\bos{G}\in\mathbb{R}^{(3N-3)\times(3N-3)}$ matrix is obtained by inversion of $\bos{g}\in\mathbb{R}^{(3N-3)\times(3N-3)}$:
\begin{align}
  \bos{G} = \bos{g}^{-1} \; .
  \label{eq:bigGmat}
\end{align}

In most applications of GENIUSH \cite{MaCzCs09,FaMaCs11,14FaMaCs,SaCsAlWaMa16,SaCsMa17,FeMa19,DaAvMa21a,DaAvMa21b}, the  discrete variable representation \cite{LiCa00} was used, and in that representation 
the Podolsky (P) form of the general vibrational KEO 
\begin{align}
\hat{H}^\text{v,P}
= 
\frac{1}{2} \sum_{k=1}^D \sum_{l=1}^D 
  \tilde{{g}}^{-1/4} 
  \hat{p}_k G_{kl} 
  \tilde{{g}}^{1/2}  
  \hat{p}_l
  \tilde{{g}}^{-1/4}  \quad\text{with}\quad \tilde g=\det\bos{g}
\end{align}
is an advantageous choice, because it requires only first-order coordinate derivatives.

In this work, we use a finite basis representation of the Hamiltonian (Secs.~\ref{ch:basis} and \ref{ch:multigrid}),
and for this purpose the `fully rearranged' form \cite{AvMa19,AvMa19b,AvPaCzMa20} is more convenient
\begin{align}
  \hat{H}^\text{v} 
  = 
  -\frac{1}{2} \sum_{i=1}^{3N-6} \sum_{j=1}^{3N-6} 
    G_{ij} \frac{\partial}{\partial q_{i}}\frac{\partial}{\partial q_{j}}
  -\frac{1}{2} \sum_{i=1}^{3N-6} B_{i}
    \frac{\partial}{\partial q_{i}}
  +U
  +V
  \label{eq:hamiltonian}
\end{align}
where $G_{ij}$, $B_i$, $U$, and $V$ are functions of the vibrational coordinates, 
\begin{align}
  B_{i} 
  = 
  \sum_{k=1}^{3N-6} 
  \frac{\partial}{\partial q_{k}} G_{ki} 
\end{align}
and the pseudo-potential term is
\begin{align}
  U= \frac{1}{32} \sum_{kl=1}^{3N-6}
    \Bigg[
    \frac{G_{kl}}{\tilde{g}^2} \frac{\partial\tilde{g}}{\partial\xi_k}
    \frac{\partial\tilde{g}}{\partial\xi_l}
    +4\frac{\partial}{\partial\xi_k} 
    \Bigg(
    \frac{G_{kl}}{\tilde{g}}
    \frac{\partial\tilde{g}}{\partial\xi_l}
    \Bigg)
    \Bigg] \; .
  \label{eq:pseudopot}
\end{align}

\section{Basis functions and truncation of the direct-product basis \label{ch:basis}}
\noindent 
We start expanding the wave function over a direct product of basis functions of the selected coordinates
\begin{align}
\Psi_{i}(q_{1},\ldots,q_{8},\tau)
=
\sum_{n_{q_{1}}=0}^{b} \ldots \sum_{n_{q_{8}}=0}^{b} 
\sum_{n_{\tor}=0}^{\nmax_{\tor}}
  C^{i}_{n_{q_{1}},\ldots,n_{q_{8}},n_{\tor}} 
  \prod_{i=1}^8 \psi^{(i)}_{n_{q_{i}}}(q_{i}) 
  \psi^{(\tor)}_{n_{\tor}}(\tor) \; .
  \label{eq:basprod}
\end{align}
In this work, we use harmonic oscillator basis functions, 
for the (dimensionless) $q_{1},\ldots,q_{8}$ \rc-normal coordinates, 
\begin{align}
  \psi_{n}(q)= A_{n} H_{n}(q)\ \mathrm{e}^{-q^2/2}
  \quad\text{with}\quad
  q\in(-\infty,+\infty) \; ,
  \label{eq:hobf}
\end{align}
where $H_{n}(q)$ is the $n$th-order Hermite polynomial and $A_{n}$ is a normalization constant.
Sec.~\ref{ch:range} provides further important technical details regarding the range of the different coordinate choices.
Regarding the $\tor$ torsional coordinate,
we use a Fourier basis including the following functions, 
\begin{align}
  1, \cos(\tor), \sin(\tor), \ldots, \cos(n_{\tor}\tor), \sin(n_{\tor}\tor) \; ,
  \quad
  \tor\in [0,2\pi) \; 
\end{align}
to solve the 1-dimensional (1D) torsional Schrödinger equation,
\begin{equation}
    \label{eq:1dhamil}
    \hat{H} = G_{\tau\tau} \frac{\partial^2}{\partial \tau^2} 
    + \frac{\partial G_{\tau\tau}}{\partial \tau} \frac{\partial}{\partial \tau} 
    + V_\tau \; ,
\end{equation}
where the pseudo-potential term, Eq.~(\ref{eq:pseudopot}), is neglected. As a result, we obtain torsional functions that can be identified as \trans, \cis, or \delocalized\ torsional functions explained and discussed in detail in Sec.~\ref{ch:numres}. We use this 1D torsional basis set to solve the 9D vibrational problem of formic acid.

An accurate product basis set representation for the lowest vibrational levels of HCOOH requires values of $\nmax_{\tor}> 30$ and $b \ge 8$. 
The number of functions in a direct product basis with $\nmax_{\tor}=30$ and $b=8$ is  $9^8\cdot 31 \approx 1.3\cdot 10^9$ that is too large for practical computations, and still not sufficient for good convergence.
Since the couplings of the eight (relaxed) curvilinear normal coordinates was made small over the entire range of $\tau$ (Sec.~\ref{ch:coor}), we may expect that the basis set can be efficiently pruned according to
\begin{align}
  \Psi_{i}
  &=
  \sum_{n_{\tor}=0}^{\nmax_{\tor}} 
  \sum_{f(n_{q_1},...,n_{q_{8}}) \le b} 
   C^{i}_{n_{q_{1}},\ldots,n_{q_{8}},n_{\tor}}
   \prod_{i=1}^8 \psi^{(i)}_{n_{q_{i}}}(q_{i}) 
  \psi^{(\tor)}_{n_{\tor}}(\tor) 
  \label{eq:basprune} \\
  &\quad\text{with}\quad  f(n_{q_1},...,n_{q_{8}}) = n_{q_1} + \ldots + n_{q_{8}} \; ,
  \label{eq:funprune}
\end{align}
where certain basis functions have been discarded from the direct product. 
Eq.~(\ref{eq:funprune}) gives the simplest possible pruning function. Poirier and co-workers \cite{HaPo15,HaPo15b,SaPo21} have studied more elaborate pruning conditions targeting very highly excited states.

Since the $\psi_{n_{q_{1}}}(q_{1})\cdot \ldots \cdot\psi_{n_{q_{8}}}(q_{8})$ product function provides a good representation for the small-amplitude (non-$\tau$) dynamics, we can discard basis functions based on simple physical arguments.
For an \emph{a priori} assessment about the importance of a basis function $|\bos{n}'\rangle$ ($\bos{n}'$ collects the basis indexes) in a wave function dominated by the $|\bos{n}\rangle$ basis state, 
the smallness of the ratio of the Hamiltonian matrix element with respect to the difference of the zeroth-order energies,
\begin{align}
\frac{%
  \langle  
    n_{1},\ldots,n_{8},n_{\tor} 
    | \hat{H} |
    n_{1}',\ldots,n_{8}',n_{\tor}' 
  \rangle
}{%
  E^{(0)}_{n_{1},\ldots,n_{8},n_{\tor}}
  -
  E^{(0)}_{n'_{1},\ldots,n'_{8}, n_{\tor}'}
}\approx 0
\label{eq:coup}
\end{align}
can provide a good indication about the unimportance of $|\bos{n}'\rangle$ for the variational result. 
The ratio is small, if 
(a) the Hamiltonian matrix element is small, and/or (b) the zeroth-order energy difference is large.
The order of magnitude of the Hamiltonian matrix element can be estimated by considering the fast convergence of the Taylor expansion of the potential and the kinetic energy in \rc-normal coordinates.

If the zeroth-order energy for a multi-dimensional basis function is very large, then the contribution of the function to the lowest-energy wave functions is negligible. 
For example, in order to compute the ground vibrational wave function, 
the 8D  basis functions 
\begin{align}
  &\bos{0}=(0,0,0,0,0,0,0,0)\quad\text{and}  \nonumber \\
  &1_1, 2_1, 3_1, 4_1, 5_1, 6_1, 7_1, 8_1 \label{eq:holist}
\end{align}
are necessary, since $\langle  \bos{0},n_{\tor} | \hat{H} | n'_{q_1},\ldots,n'_{q_8},n_{\tor}'  \rangle$ is not small. 
In Eq.~(\ref{eq:holist}), we have introduced a short notation, we list only the degrees of freedom for which the basis function index (`vibrational quantum number') is larger than 0, \emph{e.g.,} $3_1=(0,0,1,0,0,0,0,0)$.

Furthermore, less important, but still significant contribution to the ground vibrational state may be expected from the following 8D basis functions: 
\begin{align}
  &1_2,2_2, 3_2, 4_2, 5_2, 6_2, 7_2, 8_2,  \nonumber \\
  &1_12_1, 1_13_1, 1_14_1, 1_15_1, 1_16_1, 1_17_1, 1_18_1, \nonumber  \\
  &2_13_1, 2_14_1, 2_15_1, 2_16_1, 2_17_1, 2_18_1, \nonumber \\
  &3_14_1, 3_15_1, 3_16_1, 3_17_1, 3_18_1, \nonumber \\
  &4_15_1, 4_16_1, 4_17_1, 4_18_1, \nonumber \\
  &5_16_1, 5_17_1, 5_18_1, \nonumber \\
  &6_17_1, 6_18_1, \nonumber \\
  &7_18_1 \; ,
  \label{eq:alsoholist}
\end{align}
where we note that the functions with $8_1$ contribution can be discarded for the present system (HCOOH) due to symmetry reasons.
At the same time, the basis function 
$1_12_13_14_15_16_17_18_1$
gives a negligible contribution to the lowest-energy states in comparison with the  basis functions listed in Eqs.~(\ref{eq:holist}) and (\ref{eq:alsoholist}), since both the Hamiltonian matrix elements are expected to be small and the zeroth-order energy differences are large.

These kinds of arguments do not apply for discarding torsional functions, since the coupling of the $\tor$ coordinate and the curvilinear normal modes  (the Hamiltonian matrix element) may be large and the zeroth-order torsional energies are small, \emph{i.e.,} both (a)--(b) pruning `criteria' below Eq.~(\ref{eq:coup}) fail. 
Therefore, we retain all torsional basis functions in the basis set.

All in all, using the simplest pruning function in Eq.~(\ref{eq:funprune}), the direct-product basis, Eq.~(\ref{eq:basprod}), including $1.3\cdot 10^9$ functions can be reduced to $4\cdot 10^5$ functions, while the lowest (few hundred) vibrational states can be computed accurately.

For future work, we consider more elaborate pruning conditions. 
Based solely on the harmonic frequencies, the following condition could be formulated, 
\begin{align}
  0 \le 2 n_{q_{1}}+2 n_{q_{2}}+\frac{3}{2}n_{q_{3}}+n_{q_{4}}+n_{q_{5}}+n_{q_{6}}+\frac{1}{2}n_{q_{7}}+n_{q_{8}} \le b\; .
  \label{2pru}
\end{align}
This condition accounts only for the denominator of Eq.~(\ref{eq:coup}). Since the higher-frequency harmonic oscillator modes correspond to the stretching degrees of freedom, the coupling through the Hamiltonian matrix element, numerator of Eq.~(\ref{eq:coup}), may be large.  
A `safe' improvement of Eq.~(\ref{eq:funprune}), reads as
\begin{align}
  0 \le n_{q_{1}}+n_{q_{2}}+n_{q_{3}}+n_{q_{4}}+n_{q_{5}}+n_{q_{6}}+\frac{1}{2} n_{q_{7}}+ n_{q_{8}} \le b \;
  \label{3pru}
\end{align}
that corresponds to increasing the number of basis functions  for the lowest-frequency (totally symmetric) harmonic mode (O--C--O bending). We plan to use Eq.~(\ref{3pru}) in future work (see also Sec.~\ref{ch:numres}).

\section{Multi-dimensional integration\label{ch:multigrid}}
\subsubsection{Coordinate ranges for curvilinear normal coordinates \label{ch:range}} 
Since we use harmonic oscillator basis functions, Eq.~(\ref{eq:hobf}), it will be appropriate to use a set
of (nested) Hermite quadratures to evaluate integrals for the matrix elements.  Before doing that we need to address the question of the coordinate range for \rc-normal coordinates.

The range of the harmonic oscillator functions and also of the Hermite quadrature is $(-\infty,\infty)$,
whereas the range of the internal coordinates is more restricted: $[0,\infty)$ for a stretching and $[0,\pi]$ for a bending.  
At the same time, if we calculate the value of the curvilinear coordinates, $\xi_i$, using the values of \rc-normal coordinates, $\tmQ_k$ (at grid points) according to Eq.~(\ref{eq:rcnormal}), it can happen that we obtain a value that is outside
the coordinate range (\emph{e.g.,} negative value for a distance). Fortunately, this does not happen for grid points near the origin, but at the edges of the multi-dimensional grid, there are points that return internal coordinate values outside their range. 
For rectilinear normal coordinates, this does not happen, since the body-fixed Cartesian coordinates are also defined over $(-\infty,+\infty)$.

To handle the problematic points of the \rc-normal coordinate grid, we define mapping functions which ensure that the result is in the good range: 
\begin{align}
  \mathcal{F}\left[%
    \xi^{(\eq)}_i + \sum_{k=1}^D \tmL_{ik} \tmQ_k
  \right]
  \in \text{Range}(\xi_i) \;.
\end{align}
Furthermore, we expect that a good mapping function behaves as a linear (an almost trivial) mapping within the good range, but it ensures that at the `edges' of the multi-dimensional grid meaningful values are returned.

\begin{figure}
    \centering
    \includegraphics[scale=0.9]{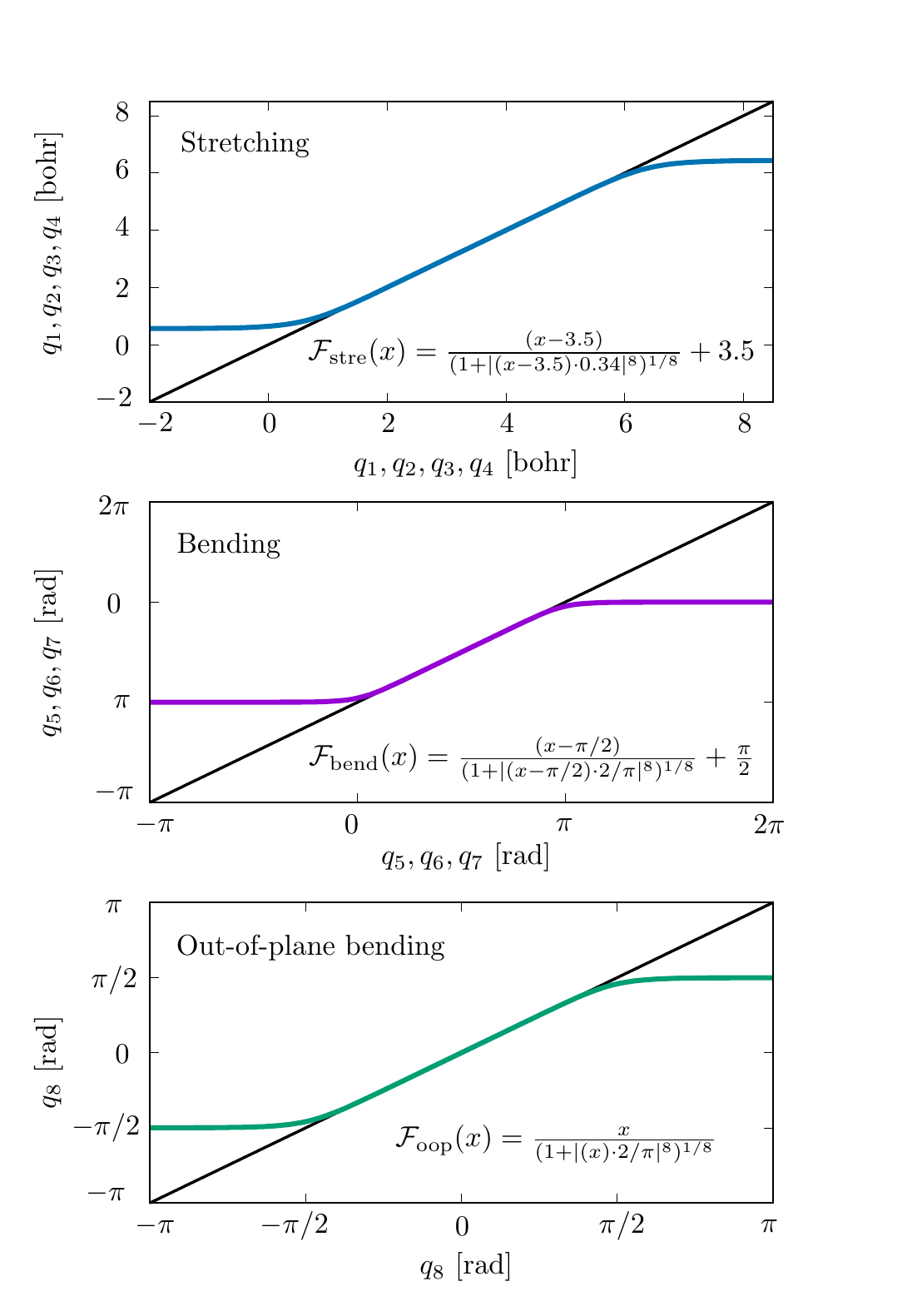}
    \caption{%
      Functions used for the stretching, bending, and out-of-plane bending coordinates to map the $(-\infty,+\infty)$ range of the curvilinear normal coordinates  to the mathematically appropriate and dynamically relevant range of HCOOH.
    }
    \label{fig:stype-fun}
\end{figure}

For a sigmoid-like mapping function, $x\mapsto\text{arctan}(x)$ and $x\mapsto\text{tanh}(x)$ are the most common examples. Unfortunately, outside the $[-0.5,0.5]$ interval, these functions significantly deviate from $x\mapsto x$, and we would like to find a function that is close to $x\mapsto x$ over almost the entire coordinate range, but ensures that the `edges' also have meaningful values. For these reasons, we decided to use 
\begin{align}
  \mF(x)= \frac{x}{(1+|x|^k)^{1/k}} 
  \label{eq:swi}
\end{align}
with $k=8$. 
The procedure is simple. For a quadrature point $\tor^{k_{\tor}}$ and $q_{1}^{k_1},\cdots,q_{8}^{k_8}$, the internal coordinates of the reference structure and the corresponding displacements are calculated. Their sum provides the `raw' internal coordinate value that is mapped to the final value of the coordinate within the correct range. 
The following parameterization is used for the stretching ($r_i$), bending ($\theta_i$), and out-of-plane bending (oop, $\varphi$) types of coordinates  (Fig.~\ref{fig:stype-fun})
\begin{align}
  \mF_\text{stre}(x)= \frac{(x-3.5)}{(1+|(x-3.5)\cdot 0.34|^8)^{1/8}} + 3.5,
\end{align}
\begin{align}
  \mF_\text{bend}(x)= \frac{(x-\pi/2)}{(1+|(x-\pi/2)\cdot 2/\pi|^8)^{1/8}} + \frac{\pi}{2},
\end{align}
and 
\begin{align}
  \mF_{\text{oop}}(x)= \frac{x}{(1+|(x)\cdot 2/\pi|^8)^{1/8}},
\end{align}
respectively.
These functions ensure that the `final' value for the stretching coordinates is within the 
$[0.56,6.44]$~bohr interval, which is the relevant dynamical range for all stretches in HCOOH, 
the value of the bending coordinate is within $[0,\pi]$, and the out-of-plane bending is within 
$[-\pi/2,\pi/2]$. Although $\varphi$ is a torsion-like coordinate and is defined on $[-\pi,\pi)$, it is a small(er) amplitude vibration of formic acid, and the relevant dynamical range is within $[-\pi/2,\pi/2]$.

Regarding Eq.~(\ref{eq:swi}), we decided to use $k=8$ because it appears to be a good compromise between a faithful mapping (of the good range) and numerical integrability of the matrix elements with a reasonable number of points.
Nevertheless, we have checked values up to $k=14$ using a pruned basis set with $b=8$ (Sec.~\ref{ch:basis}), and we obtained the vibrational energies within 0.03~\cm\ from the $k=8$ values (using the same basis) up to 3000~\cm\ beyond the zero-point energy.

\subsection{Smolyak quadrature \label{ch:grid}} 
We use the Smolyak approach \cite{Smolyak1963,AvCa09,AvCa11,AvCa11b,LaNa14,ChLa21} to construct efficient non-product grids for the pruned basis set, Eqs.~(\ref{eq:basprune})--(\ref{eq:funprune}), that can be used to evaluate the multi-dimensional integrals appearing in the kinetic energy coefficients, Eq.~(\ref{eq:hamiltonian}), and in the PES.
The Smolyak quadrature for nine dimensions is defined as
\begin{align}
  &Q(9,H)
  =
  \sum_{\bos{\sigma}_{\bos{s}}(i)\le H}
  \otimes\prod_{\chi=1}^9 \Delta \hat{Q}^{i_{q_\chi}}_\chi \; , 
  \quad
  i_{\chi}=1,2,3,4,\ldots, \chi=1,\ldots,9 \nonumber \\
  &\quad\quad\text{with}\  {{\bos{\sigma}}_{\bos{s}}}(i)=s^{\tor}(i_{\tor})+s^{q_{1}}(i_{q_{1}})+\ldots+s^{q_{8}}(i_{q_{8}}) \; ,
  \label{smolgridor}
\end{align}
where $H$ is a grid-pruning parameter, ${\bos{\sigma}_{\bos{s}}}(i)$ is a grid-pruning function, 
and the incremental operator is defined as
\begin{align}
  \Delta \hat{Q}^{i_{\chi}}_\chi= \hat{Q}^{i_{\chi}}_\chi-\hat{Q}^{i_{\chi}-1}_\chi 
\end{align}
with $\hat{Q}^{0}_\chi=0$ and 
the 1D quadrature rules, 
\begin{align}
  \hat{Q}^{i_{\chi}}_\chi f(q_{\chi})=\sum_{m=1}^{m_{i_{\chi}}} w^{i_{\chi}}_{\chi,m} f(q_{\chi,m}^{i_{\chi}}) \; ,
  \quad
  i_{\chi}=1,2,3,4,\ldots 
\end{align}
Equivalently, we can also write the Smolyak quadrature as a linear combination of product quadratures with different 1-dimensional accuracies as
\begin{align}
  &Q(9,H)
  =
  \sum_{{{\bos{\sigma}}_{\bos{s}}}(i)\le H} 
    C_{\bos{i}} 
    \left(%
      \otimes \prod_{\chi=1}^9 \hat{Q}^{i_{\chi}}_\chi 
    \right)\; ,\ i_{\chi}=1,2,3,4,\ldots, \chi=1,\ldots,9, \\
  &\quad\quad\text{with}\quad  {{\bos{\sigma}}_{\bos{s}}}(i)=s^{\tor}(i_{\tor})+s^{q_{1}}(i_{q_{1}})+\ldots+s^{q_{8}}(i_{q_{8}}) \; .
\end{align}
$Q(9,H)$  has a smaller number of points, than the direct product grid, 
$\hat{Q}^{i^{\rm max}_{q_{1}}}_{q_{1}} \otimes \ldots \otimes 
\hat{Q}^{i^{\rm max}_{q_{8}}}_{q_{8}} \otimes \hat{Q}^{i^{\rm max}_{\tor}}_{\tor} $, 
and its accuracy depends on three factors, 
(a) the form of the $s^{\chi}(i_{\chi})$ grid pruning functions, for which $s^{\chi}(i_{\chi})\ge s^{\chi}(i_{\chi}-1)$ must hold;
(b) the grid-pruning parameter $H$; and
(c) the number of the $m_{i_{\chi}}$ grid points, for which $m_{i_{\chi}}\ge m_{i_{\chi}-1}$ must hold.

For constructing the Smolyak grid in the present work, we define the $s^{\chi}(i_{\chi})$ functions as follows:
\begin{align}
  \chi=\tau:\quad &s^{\chi}(i_{\chi})=10 \\
  \chi=q_1,\ldots,q_8:\quad &s^{\chi}(i_{\chi})=10 i_\chi \;,\quad i_\chi=1,2,3,\ldots
\end{align}
The value of $H$, which sets an upper limit on the sum of the $s$ function values, is chosen according to this definition.
The sequences of quadrature rules $\hat{Q}^{i_{\chi}}_\chi$ are chosen as
\begin{align}
  \chi=\tau:& \quad
  \hat{Q}^{i_{\tau}}_\tau
  =
  \hat{Q}_{M_{\tau}^{\rm max}}^\text{trap} 
  \; ,\quad i_\tau=1,2,3,\ldots~ \nonumber \\
  \chi=q_1,\ldots,q_8:& \quad 
  \hat{Q}^{i_{\chi}}_\chi
  =
  \hat{Q}_{m_{i_{\chi}}}^{\rm Her}\; ,\quad i_\chi=1,2,3,\ldots~  \nonumber \\
  &\text{with}\quad m_{i_{\chi}}= 1,3,3,7, 9, 9, 9, 9,17,19,19,19,31,33,41,41,\ldots 
  \label{thems}
\end{align}
where $\hat{Q}_{M^{\rm max}_\tau}^\text{trap}$ is a trapezoidal quadrature rule of $M^{\rm max}_\tau$ points and a maximum degree of $d_\tau=2 M^{\rm max}_\tau-1$, 
while $\hat{Q}_{m_{i_{\chi}}}^{\rm Her}$ are nested quadrature rules for Hermite polynomials with a maximum degree of $d_{i_\chi}=$ 1, 5, 5, 7, 15, 15, 15, 15, 17, 29, 29, 29, 31, 33, 61, 61$, \ldots$ corresponding to $i_\chi=1,2,3,\ldots$ \cite{HeWi08}
Nesting means that all quadrature points of the quadrature rule $\hat{Q}^{j}$ appear in the higher-order quadrature rule, $\hat{Q}^{j+1}$.

Using this construct with $M_{\tor}=11$ trapezoidal points and $H=150$, we can integrate exactly all overlap matrix elements for the pruned basis set
with $0 \le n_{\tor} \le 4$ and $0 \le n_{q_{1}}+\ldots+n_{q_{8}} \le 8$ conditions. 
For $H=170$, the Smolyak grid includes $1\ 230\ 251\approx 1.2\cdot 10^6$ points.  
The smallest 9D direct-product Gauss grid that integrates correctly the same overlap matrix would have  $11\cdot 9^8 \approx 4.7\cdot 10^8$ points.

The Smolyak algorithm using nested sequences of quadrature rules allows us to use a non-product grid that has a structure, \emph{i.e.,} a multi-dimensional integral of a function $F(x_{1},\ldots,x_{9})$ can be written as
\begin{align}
  \int F(x_{1},\cdots,x_{9}) \dd x_{1}\ldots \dd x_{9} 
  \approx 
  \sum_{k_{1}=1}^{k_{1}^{\rm max}}\ldots\sum_{k_{9}=1}^{k_{12}^{\rm max}} 
    W^{\rm smol}(k_{1},\ldots,k_{9}) 
    F(x_{1}^{k_{1}},\ldots, x_{9}^{k_{9}}) \; ,
\end{align}
where $W^{\rm Smolyak}(k_{1},\cdots,k_{9})$ is the multi-dimensional Smolyak weight and the points are sorted according to the sequence of quadrature rules. The structure appears in the 
$k_{c}^{\rm max}$ indexes. $k_{1}$ depends on $H$, $k_{2}$ depends on $H$ and $k_{1}$, etc. and thus, matrix-vector products can be computed by sequential summation \cite{BraCa94,AvCa11b,AvCa09,WaCa03,CaWa11,AvMa19,AvMa19b,AvPaCzMa20}. 
Eigenvalues and eigenvectors are computed using a Lanczos iterative eigensolver that requires only the multiplication of the Hamiltonian matrix with a vector. Implementation details regarding the matrix-vector multiplication has been described in Refs.~\cite{AvCa09,AvCa11b,AvMa19}.

\section{Numerical results \label{ch:numres}} 
\noindent We have computed the vibrational energies using the basis set and pruning condition defined in Eqs.~(\ref{eq:basprune}) and (\ref{eq:funprune}). 
The number of torsional basis functions was 55 and we used 79 grid points for this degree of freedom.
Regarding the 8-dimensional (8D) \rc-normal coordinate part of the problem, three basis set sizes were used with the $b=8,$ 9, and 10 basis pruning parameter and with the $H=190$, 200, and 210 grid pruning parameter, respectively. 
As a result, the 9D basis sets with $b=8,9,$ and 10 included $707\ 850$, $1\ 337\ 050$, and $2\ 406\ 690$ basis functions, respectively. The size of the corresponding non-product Smolyak quadrature grid was $42\ 223\ 623$, $72\ 656\ 063$ and $132\ 043\ 839$. 

\begin{figure}
    \centering
    \includegraphics[scale=0.8]{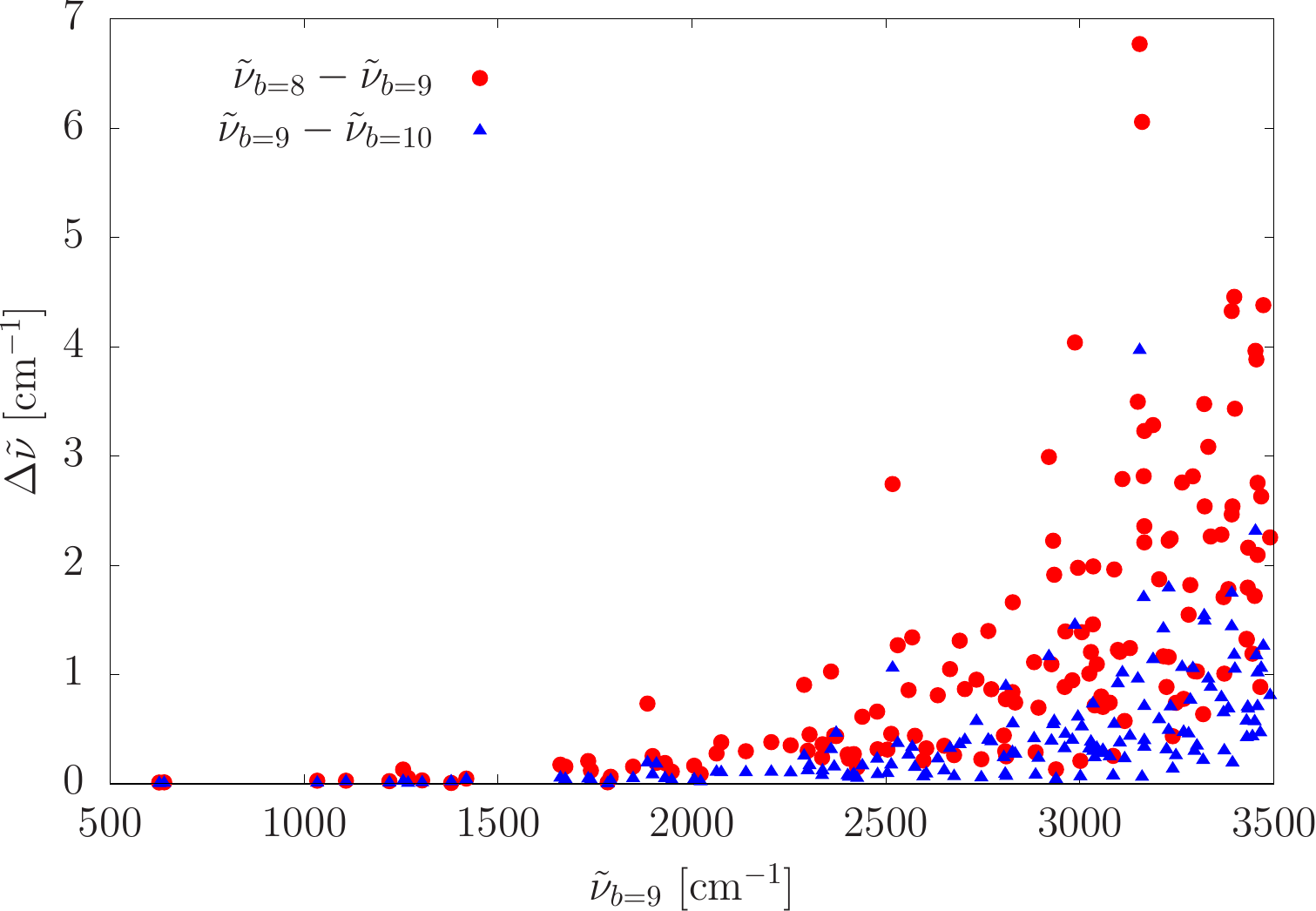}
    \caption{
    Assessment of the convergence of the vibrational band origin obtained with a $b=9$ pruned basis set, Eqs.~(\ref{eq:basprune})--(\ref{eq:funprune}), using relaxed curvilinear (\rc) normal coordinates for the non-torsional degrees of freedom. The vibrational energies are compared with smaller ($b=8$) and larger ($b=10$) basis-set results. 
    The zero-point energy with $b=8,9,$ and 10 is $\tilde\nu_0=7350.84,7350.82,$ and 7350.81~\cm, respectively. 
    }
    \label{fig:conv}
\end{figure}

The value of $H$ was selected to be able to exactly integrate
the Hamiltonian matrix elements up to 5th order (in a hypothetical Taylor expansion) 
with the highest-excited basis functions in the pruned basis set.
Of course, we have checked the effect of using a larger $H$ value. For $H=200$ (instead of $H=190$) with $b=8$, the eigenvalues up to 5000 cm$^{-1}$, beyond the zero-point vibrational energy (ZPVE), changed at most by $0.001$~cm$^{-1}$. 
Based on these observations, we think that the procedure is almost perfectly variational, which corresponds to an exact integration and provides rigorous energy upper bounds.
Further computations with the more elaborate pruning condition in Eq.~(\ref{3pru}), with larger basis and grid sizes are in progress and will allow us to have access to well-converged vibrational energies beyond 2000~\cm\ above the ZPVE.

\begin{figure}
    \centering
    \includegraphics[scale=0.8]{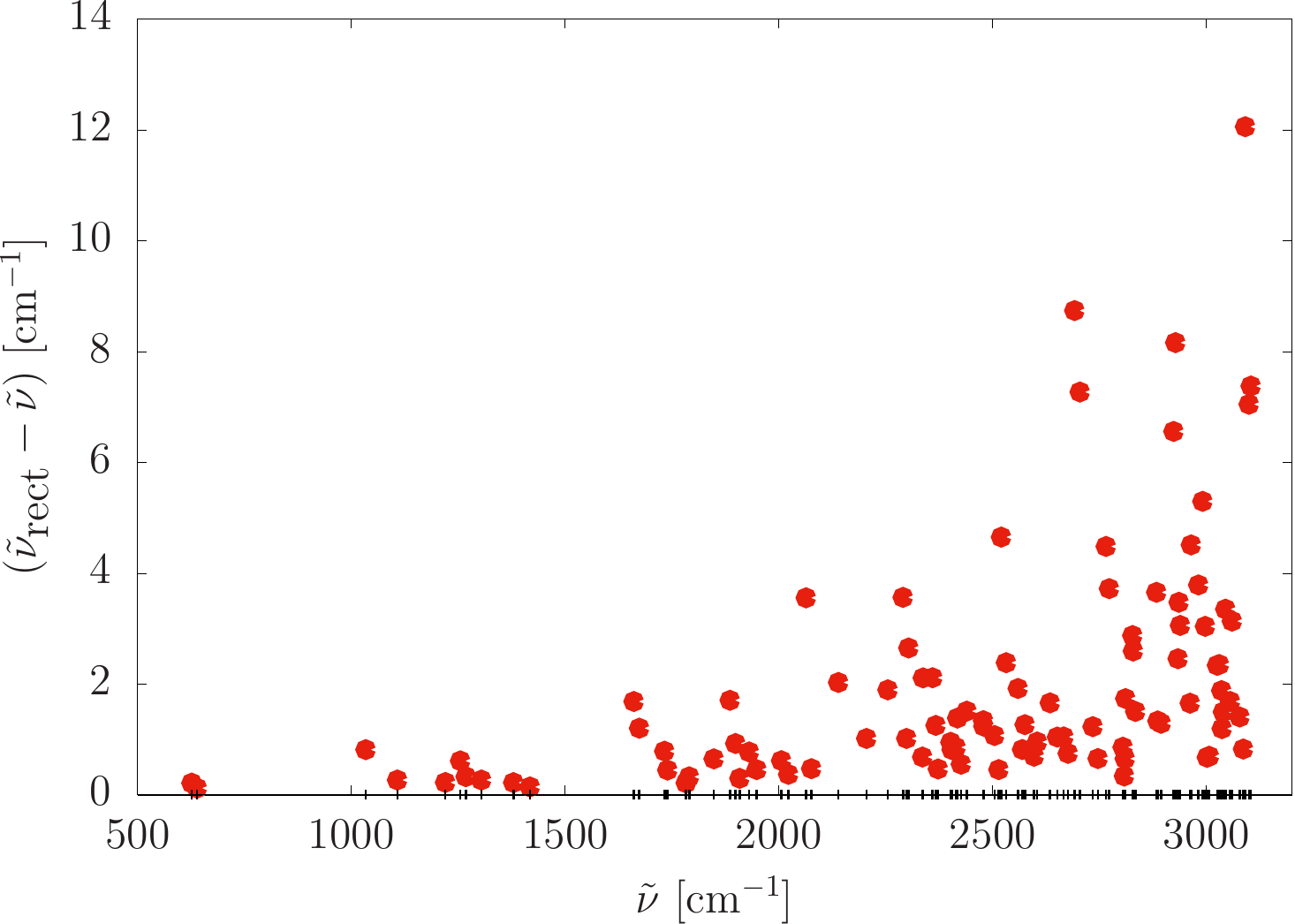}
    \caption{%
    Difference between the vibrational energies obtained using rectilinear, $\tilde\nu_\text{rect}$, and relaxed curvilinear (\rc), $\tilde\nu$, normal coordinates for the non-torsional degrees of freedom.
    In both cases the pruned basis set corresponds to the $b=8$ parameter, Eqs.~(\ref{eq:basprune})--(\ref{eq:funprune}). 
    The corresponding ZPVEs are $\tilde\nu_\text{0}=7350.84$~\cm\ and $\tilde\nu_\text{rect,0}=7350.91$~\cm.
    }
    \label{fig:curv_energy}
\end{figure}

Convergence properties have been tested with respect to the basis set size (Fig.~\ref{fig:conv}) and the coordinate representation (Fig.~\ref{fig:curv_energy}). Figure~\ref{fig:conv} shows that the vibrational energies with $b=9$ are converged better than 1~\cm\ up to ca. 2500~\cm, and within ca. 2~\cm\ up to 3500~\cm\ beyond the ZPVE.

Figure~\ref{fig:curv_energy} highlights the efficiency of the \rc-normal coordinates (Sec.~\ref{ch:coor}) in comparison with the rectilinear normal coordinates (for the non-torsional degrees of freedom). For the $b=8$ basis, the rectilinear normal coordinate vibrational energies differ (are less accurate) by 8--10~\cm\ from the \rc-normal coordinate results. 

We have also tested the coordinate definition of Lauvergnat and Nauts who computed the  vibrational states of the methanol molecule in full dimensionality (12D) \cite{LaNa14}. They did not relax and interpolate the normal coordinate coefficients along the large-amplitude coordinate, but they calculated the average of the (curvilinear) normal coordinate coefficients at the two local minima, hence we may call their coordinates averaged, curvilinear (\ac) normal coordinates. For HCOOH, we have performed computations both with \ac- and \rc-normal coordinates with the $b=8$ basis set. We have found that (for the present system) the relaxed curvilinear (\rc) normal coordinates slightly outperform simple averaging (\ac-normal coordinates), but the difference is typically less than 0.5~\cm\ in the higher energy range. Hence, the \ac-normal coordinates appear to be an excellent choice and they are technically much simpler to construct than the \rc-normal coordinates used in the present work. Nevertheless, if there are multiple minima, stronger coupling of the large-amplitude motion with the `rest' of the molecule, or more than one large-amplitude motions, then we can anticipate that the relaxation-interpolation approach used in the present work is, in principle, more efficient.

\begin{table}[]
    \centering
    \caption{%
      Plane reflection symmetry ($C_s$ point group) with respect to the equilibrium structures of the one-dimensional basis functions used in the computations (Sec.~\ref{ch:basis}).  The first 22 1-dimensional (1D) torsional functions ($c,t,d$) are plotted in Fig.~\ref{fig:cistrans}. 
      ($n=0,1,2,\ldots,$ $m=1,2,\ldots$)
      \label{tab:planesym}
    }
    \begin{tabular}{@{}l @{\ \ \ \ \ \ } l @{}}
      \hline\hline\\[-0.35cm]
         A$'$     &    A$''$ \\
      \hline\\[-0.35cm]         
         $1_n$, $2_n$, $3_n$, $4_n$, $5_n$, $6_n$, $7_n$ &
         -- \\
         $8_0$, $8_2$, \ldots, $8_{2n}$ &
         $8_1$, $8_3$, \ldots, $8_{2n+1}$ \\
         $t_0$, $t_2$, $t_4$, $t_6$ & 
         $t_1$, $t_3$, $t_5$, $t_7$\\
         $c_0$, $c_2$, $c_4$, $c_6$ & 
         $c_1$, $c_3$, $c_5$ \\
         $d_0$, $d_3$, $d_4$, \ldots, $d_{4m-1}$, $d_{4m}$ &
         $d_1$, $d_2$, $d_5$, $d_6$, \ldots, $d_{4m-3}$, $d_{4m-2}$ \\
         %
         %
      \hline\hline
    \end{tabular}
\end{table}

\subsection{Torsional assignment}
\begin{figure}
    \centering
    \includegraphics[scale=1.]{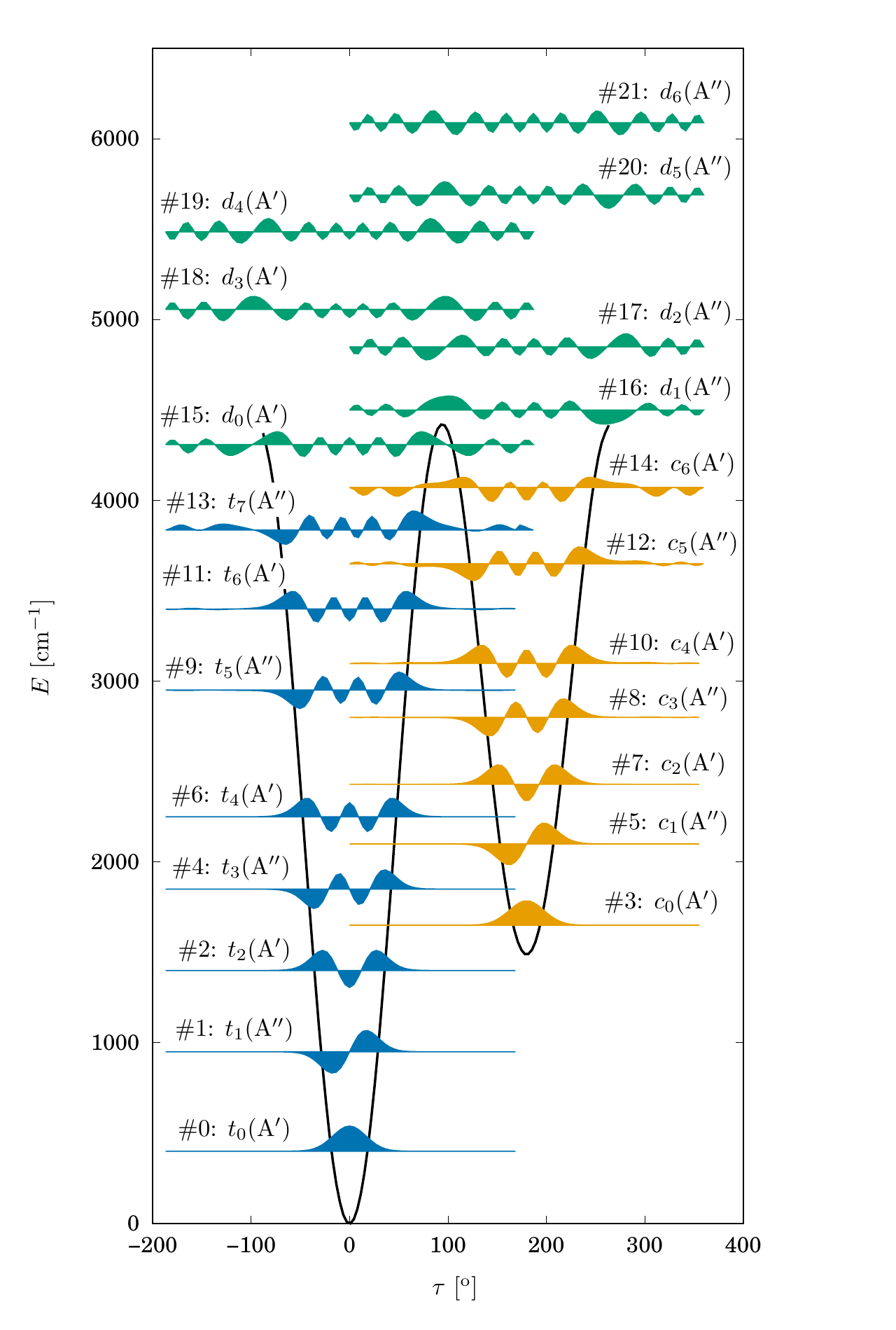}
    \caption{%
      Visualization of the \trans- (blue), the \cis- (yellow), and the lowest-energy \emph{delocalized} (green) 1D torsional functions that were obtained by solving the 1D torsional Schrödinger equation, Eq.~(\ref{eq:1dhamil}). 
      The symmetry properties of the functions with respect to reflection to the plane defined by the equilibrium structures ($\tau=0^\text{o}$ and 180$^\text{o}$ in the figure) are collected in Table~\ref{tab:planesym}.
    }
    \label{fig:cistrans}
\end{figure}

Since the computation is not localized to one of the wells of the PES (Fig.~\ref{fig:cistrans}), it is a relevant question to ask whether a given state can be assigned to the \trans\ or the \cis\ conformer. 
The torsional assignment of the 9D wave functions was performed based on the contribution of the 1D torsional basis functions (Fig.~\ref{fig:cistrans}). 
Unless the torsional energy is very high, the torsional functions are localized in the \trans\ or in the \cis\ well, \emph{i.e.,} they have a well-defined number of nodes beyond the ground state in `their' well. 
The 1D torsional functions are eigenfunctions of the Schrödinger equation with the 1D torsional Hamiltonian, Eq.~(\ref{eq:1dhamil}).
There are eight 1D \trans\ torsional functions ($t_0,t_1,\ldots,t_7$) and there are seven 1D \cis\ torsional functions ($c_0,c_1,\ldots,c_6$). Beyond these states, the torsional functions 
have nodes in both wells and we call them \emph{delocalized} functions. Each torsional function has a well-defined parity with respect to reflection to the plane of the equilibrium structures (Table~\ref{tab:planesym}). 
The torsional assignment of a 9D vibrational state was performed based on the assignment of the dominant torsional function. The plane reflection symmetry of the 9D vibrational wave function can be determined by the symmetry of the torsional functions and the symmetry of the out-of-plane vibrational mode (Table~\ref{tab:planesym}).

\begin{figure}
  \centering
  \includegraphics[scale=0.8]{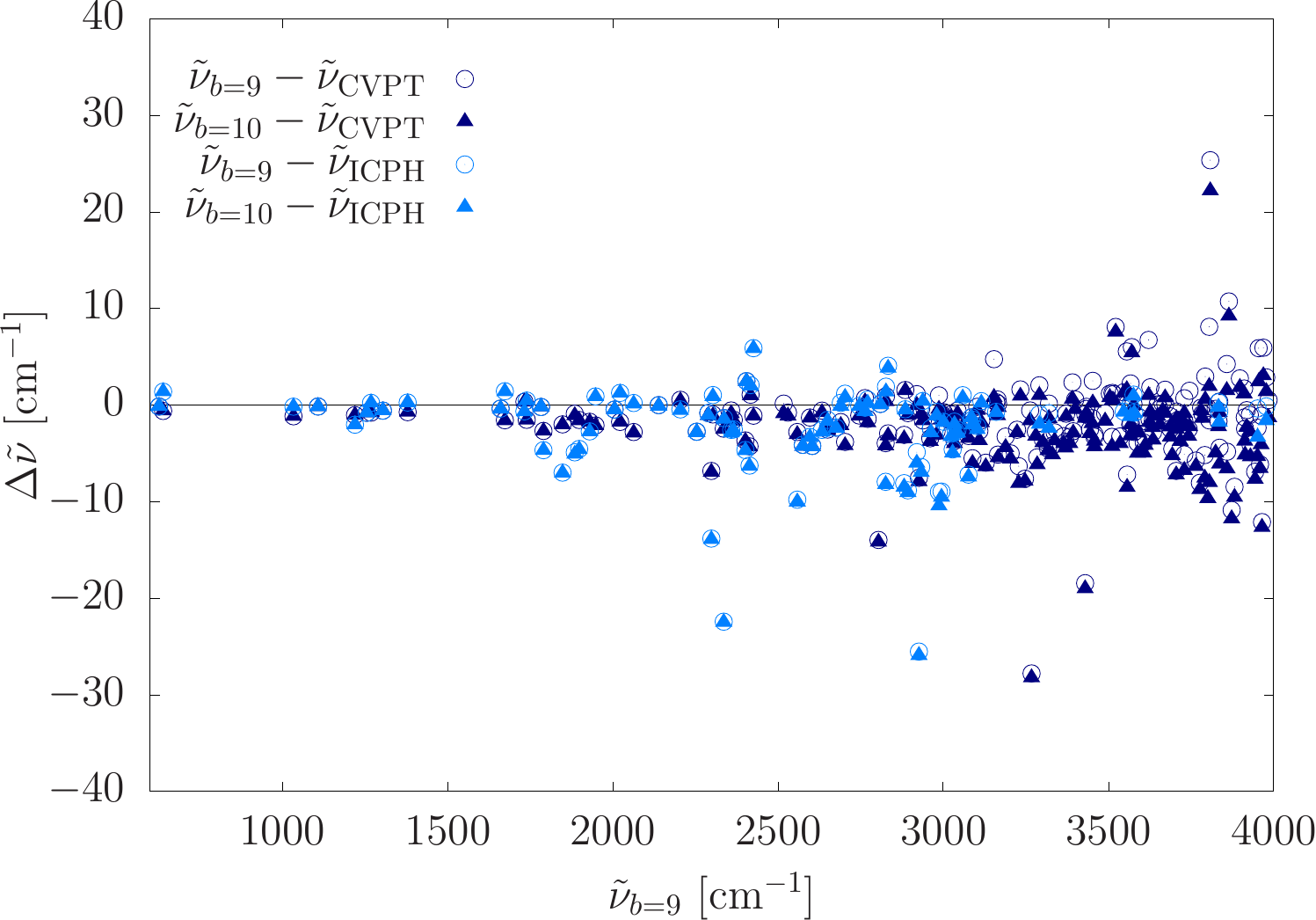}
  \caption{%
    Comparison of \trans-HCOOH vibrational energies, $\tilde\nu$ referenced to the zero-point energy, computed with the \gesm\ approach using the $b=9$ and $10$ basis sets, Eqs.~(\ref{eq:basprune}) and (\ref{eq:funprune}), and the 6th-order canonical van Vleck perturbation theory (CVPT) \cite{NeSi21} and the internal-coordinate path Hamiltonian (ICPH) \cite{TeMi16} results.
    The states from the different computations were compared based on the assignment of their wave function (Tables~\ref{tab:trans1}--\ref{tab:trans3}).
    The zero-point energies are  $\tilde\nu_{\text{ZPV}}=7351$~\cm\ ($b=9$ and 10) and
    $\tilde\nu_{\text{ICPH},\text{ZPV}}=7354$~\cm. 
    \label{fig:transdiff}
  }
\end{figure}

\begin{figure}
  \centering
  \includegraphics[scale=0.8]{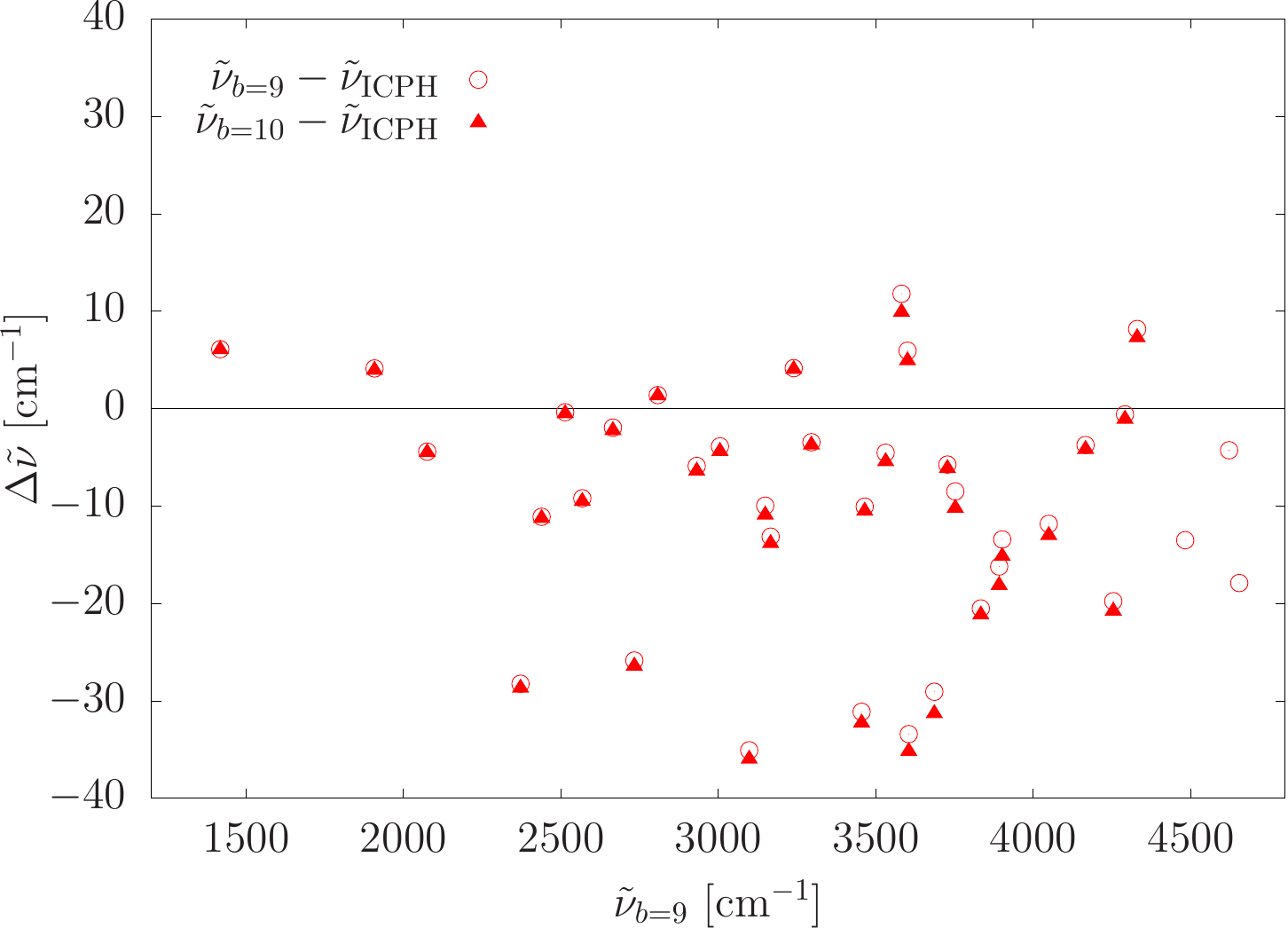}
  \caption{%
    Comparison of the \cis-HCOOH vibrational energies, $\tilde\nu$ referenced to the zero-point energy, computed with the \gesm\ approach using the $b=9$ and $10$ basis sets, Eqs.~(\ref{eq:basprune}) and (\ref{eq:funprune}), and the internal-coordinate path Hamiltonian (ICPH) \cite{TeMi16} results.
    The states from the different computations were compared based on the assignment of their wave function (Tables~\ref{tab:cis1} and \ref{tab:cis2}).
    The (\trans-)zero-point energies are $\tilde\nu_{\text{ZPV}}=7351$~\cm\ ($b=9$ and 10) and $\tilde\nu_{\text{ICPH},\text{ZPV}}=7354$~\cm. 
    \label{fig:cisdiff}
  }
\end{figure}

\clearpage

Regarding the \trans\ states computed with the \gesm\ approach ($b=9$ and 10), we observe an overall good agreement with the internal-coordinate path Hamiltonian (ICPH) \cite{TeMi16} and a very good agreement with the 6th-order canonical van Vleck perturbation theory (CVPT) \cite{NeSi21} results (Fig.~\ref{fig:transdiff}).
The present variational computations systematically improve upon the CVPT results by 5--10(--25)~\cm\ up to 4000~\cm\ beyond the ZPVE. The good agreement of the \trans\ vibrational energies (and assignments) with CVPT is interesting, since the CVPT computation was based on a single-well description and the \cis\ zero-point vibration is only 1418~\cm\ higher than the \trans\ ZPVE, but it can be explained by the relatively high \cis-\trans\ isomerization barrier (Fig.~\ref{fig:cistrans}).

The current (almost perfectly variational) computation improves the CVPT results by 1--5~\cm\ in the range up to ca. 2500~\cm, and by ca. 5--15~\cm\ in the 2500--4000~\cm\ range.
We can spot one important outlier from this favorable comparison at around 3808~\cm\ (Fig.~\ref{fig:transdiff}). For this state the \gesm\ ($b=9$) energy is by 25~\cm\ \emph{higher} than the CVPT energy, and by comparing the $b=9$ and $10$ energies, it is unlikely that some further enlargement of the basis set ($b=11,$ 12) reduces this deviation to a value below 5~\cm.
This state is unambiguously assigned to $7_6t_0$ in both computation, which is the 6th excitation of the lowest-energy, totally symmetric harmonic mode ($\nu_7$). By considering the currently used pruning condition, Eq.~(\ref{eq:funprune}), and the $b=8,9,10$ (11, 12) values of the pruning parameters (that are computationally feasible), we think that for this state our result is in an error and CVPT makes a (probably) good prediction. 
We can improve the \gesm\ results, without significantly increasing the computational cost, 
by using the more elaborate pruning condition in Eq.~(\ref{3pru}). The Eq.~(\ref{3pru}) pruning simply allows to double the number of basis functions for the totally symmetric,  lowest-frequency mode, while keeping the size of the multi-dimensional basis within reasonable limits. Further work in this direction is in progress and will be reported in the future.

Regarding the \cis\ states, the computed fundamentals agree well from all computations (\gesm, ICPH, and CVPT). We note that the highest energy $c_1$ vibration is not (yet) available from the \gesm\ computations (this work), nor from ICPH Ref.~\cite{TeMi16}. Although CVPT results are available for all \cis\ fundamental vibrations \cite{NeSi21}, combination and overtone bands have not been reported. 

Figure~\ref{fig:cisdiff} shows the comparison for \cis\ states up to two excitations (based on the wave function assignments, Tables~\ref{tab:cis1}--\ref{tab:cis2}) of the \gesm\ ($b=9,$ 10) and the ICPH \cite{TeMi16} results. Similarly to the \trans\ energies, there is an overall good agreement, but several ICPH energies are too high (by 10--40~\cm). 

Beyond 3700~\cm\ (above the \trans-ZPVE), we can see mixed \cis-\trans\ states, in which torsional functions corresponding to the \trans\ and other functions corresponding to the \cis\ well are entangled (Tables~\ref{tab:cis1}--\ref{tab:cis2}). 
Several \cis-\trans\ entangled states come in pairs corresponding to $+$ and $-$ combinations of \cis\ and \trans\ basis functions. The corresponding `tunneling splittings', which we currently compute to be $<1-5(-10)$~\cm, are smaller or on the borderline with respect to the convergence uncertainty of the $b=9$ basis in the relevant energy range (indicated by $(\delta_{10},\delta_8)$ in the tables), and we plan to determine these splittings more precisely in future work.

Beyond 3900~\cm\ (above the \trans-ZPVE), non-negligible contribution from delocalized torsional states (Fig.~\ref{fig:cistrans}) can be observed. Table~\ref{tab:deloc} shows the lowest-energy  vibrational states with a significant delocalized contribution (see also Tables~\ref{tab:cis1}--\ref{tab:cis2}). These states have an energy close to the \cis-\trans\ isomerization barrier height (Fig.~\ref{fig:cistrans}) and in this range the \cis-\trans-\emph{delocalized} functions strongly mix
and for their good description a variational procedure appears to be necessary. Further, better converged results will be reported in future work.

\begin{table}[H]
\singlespacing
  \caption{%
    \emph{Trans}-HCOOH: vibrational excitation energies, $\tilde\nu$ in \cm,
    computed with the \gesm\ approach in comparison with the
    sixth-order canonical van Vleck perturbation theory (CVPT) \cite{NeSi21} and
    the internal-coordinate reaction-path Hamiltonian (ICPH) \cite{TeMi16} results.
    (The footnotes are below Table~\ref{tab:trans3}.) 
    \label{tab:trans1}
  } 
  \centering
\vspace{-0.4cm}  
\singlespacing
\scalebox{1.}{
  \begin{tabular}{@{}c l c@{\ }ccc@{\ \ \ \ } c l c@{\ }ccc@{}}
    \hline\hline\\[-0.35cm]
    \#$^\text{a}$ &	
    Assign.$^\text{b}$ & 
    $\tilde\nu$ &	
    $(\delta_{10},\delta_8)^\text{c}$	& 
    CVPT &
    ICPH  &
    \#$^\text{a}$ &	
    Assign.$^\text{b}$ & 
    $\tilde\nu$ &	
    $(\delta_{10},\delta_8)^\text{c}$	& 
    CVPT  &
    ICPH  \\
    \hline\\[-0.35cm]
1	&	ZPV-$t_0$	&	7351	&	(0,0)	&	[n.a.]	&	7354	&	71	&	$6_27_1t_0$	&	2827	&	(1,2)	&	2826	&	2825	\\
2	&	$7_1t_0$	&	627	&	(0,0)	&	627	&	627	&	72	&	$6_2t_1$	&	2833	&	(0,1)	&	2836	&	2829	\\
3	&	$t_1$	&	639	&	(0,0)	&	640	&	638	&	73	&	$7_18_1t_2$	&	2882	&	(0,1)	&	2885	&	2890	\\
4	&	$8_1t_0$	&	1034	&	(0,0)	&	1035	&	1034	&	74	&	$3_16_1t_0$	&	2886	&	(0,0)	&	2884	&	2886	\\
5	&	$6_1t_0$	&	1108	&	(0,0)	&	1108	&	1108	&	75	&	$6_1t_3$	&	2893	&	(0,1)	&	2894	&	2902	\\
6	&	$t_2$	&	1220	&	(0,0)	&	1221	&	1222	&	76	&	$7_38_1t_0$	&	2920	&	(1,3)	&	2919	&	2925	\\
7	&	$7_2t_0$	&	1255	&	(0,0)	&	1256	&	1256	&	77	&	[$7_1t_4$,$6_17_1t_2$]	&	2926	&	(0,1)	&	2934	&	2952	\\
8	&	$7_1t_1$	&	1268	&	(0,0)	&	1269	&	1268	&	79	&	$7_28_1t_1$	&	2934	&	(1,2)	&	2934	&	2940	\\
9	&	$5_1t_0$	&	1304	&	(0,0)	&	1305	&	1305	&	80	&	$2_1t_0$	&	2938	&	(0,0)	&	2940	&	2938	\\
10	&	$4_1t_0$	&	1379	&	(0,0)	&	1380	&	1379	&	81	&	[$7_1t_4$,$6_17_1t_2$]	&	2961	&	(0,1)	&	2964	&	3066	\\
12	&	$7_18_1t_0$	&	1661	&	(0,0)	&	1661	&	1661	&	82	&	$5_17_18_1t_0$	&	2963	&	(0,1)	&	2965	&	2965	\\
13	&	$8_1t_1$	&	1673	&	(0,0)	&	1675	&	1672	&	83	&	$5_18_1t_1$	&	2980	&	(0,1)	&	2983	&	2981	\\
14	&	$6_17_1t_0$	&	1732	&	(0,0)	&	1732	&	1733	&	84	&	$6_17_3t_0$	&	2987	&	(1,4)	&	2986	&	2996	\\
15	&	$6_1t_1$	&	1739	&	(0,0)	&	1741	&	1739	&	85	&	$6_17_2t_1$	&	2995	&	(1,2)	&	2996	&	3004	\\
16	&	$3_1t_0$	&	1783	&	(0,0)	&	1783	&	1783	&	86	&	$3_1t_2$	&	3001	&	(0,0)	&	3002	&	3003	\\
17	&	$t_3$	&	1790	&	(0,0)	&	1793	&	1795	&	88	&	$3_17_2t_0$	&	3024	&	(0,1)	&	3025	&	3027	\\
18	&	$7_1t_2$	&	1848	&	(0,0)	&	1850	&	1855	&	89	&	[$5_16_17_1t_0$]	&	3028	&	(0,1)	&	3030	&	3033	\\
19	&	$7_3t_0$	&	1885	&	(0,1)	&	1886	&	1890	&	91	&	[$5_16_1t_1$]	&	3034	&	(0,2)	&	3038	&	3036	\\
20	&	$7_2t_1$	&	1898	&	(0,0)	&	1900	&	1903	&	92	&	$3_17_1t_1$	&	3038	&	(0,1)	&	3040	&	3041	\\
22	&	$5_17_1t_0$	&	1930	&	(0,0)	&	1932	&	1933	&	93	&	$4_17_18_1t_0$	&	3043	&	(0,1)	&	3044	&	3046	\\
23	&	$5_1t_1$	&	1948	&	(0,0)	&	1950	&	1947	&	94	&	[$7_1t_1$,$7_2t_3$]	&	3055	&	(0,1)	&	3057	&	[3109]	\\
24	&	$4_17_1t_0$	&	2006	&	(0,0)	&	2006	&	2006	&	95	&	$4_18_1t_1$	&	3059	&	(0,1)	&	3061	&	3058	\\
25	&	$4_1t_1$	&	2022	&	(0,0)	&	2024	&	2021	&	96	&	[$5_1t_3$]	&	3077	&	(0,1)	&	3080	&	3084	\\
26	&	$8_2t_0$	&	2063	&	(0,0)	&	2066	&	2063	&	97	&	$3_15_1t_0$	&	3086	&	(0,0)	&	3087	&	3087	\\
28	&	$6_18_1t_0$	&	2139	&	(0,0)	&	2139	&	2139	&	98	&	$8_3t_0$	&	3089	&	(1,2)	&	3094	&	3090	\\
29	&	$6_2t_0$	&	2205	&	(0,0)	&	2204	&	2205	&	100	&	$4_16_17_1t_0$	&	3103	&	(0,1)	&	3103	&	3105	\\
30	&	$8_1t_2$	&	2254	&	(0,0)	&	2257	&	2257	&	101	&	[$7_3t_2$]	&	3109	&	(1,3)	&	3112	&	[3144]	\\
31	&	$7_28_1t_0$	&	2289	&	(0,1)	&	2290	&	2290	&	102	&	$4_16_1t_1$	&	3115	&	(0,1)	&	3117	&	3115	\\
32	&	[$t_4$,$6_1t_2$,$5_1t_2$]	&	2298	&	(0,0)	&	2305	&	2312	&	103	&	[$5_27_1t_0$]	&	3129	&	(0,1)	&	3135	&	[3159]	\\
33	&	$7_18_1t_1$	&	2303	&	(0,0)	&	2304	&	2302	&	105	&	$7_5t_0$	&	3154	&	(4,7)	&	3149	&		\\
34	&	[$t_4$,$6_1t_2$]	&	2336	&	(0,0)	&	2338	&	2358	&	106	&	$3_14_1t_0$	&	3160	&	(0,6)	&	3160	&	3161	\\
35	&	$5_18_1t_0$	&	2337	&	(0,0)	&	2338	&	2338	&	107	&	$7_4t_1$	&	3165	&	(2,3)	&	3164	&		\\
36	&	$6_17_2t_0$	&	2358	&	(0,1)	&	2359	&	2361	&	109	&	$4_1t_3$	&	3166	&	(0,2)	&	3171	&		\\
37	&	$6_17_1t_1$	&	2366	&	(0,0)	&	2368	&	2369	&	110	&	$6_18_2t_0$	&	3166	&	(0,3)	&	3166	&		\\
39	&	[$5_16_1t_0$]	&	2401	&	(0,0)	&	2405	&	2406	&	111	&	$5_17_3t_0$	&	3189	&	(1,3)	&	3192	&		\\
40	&	$3_17_1t_0$	&	2404	&	(0,0)	&	2402	&	2402	&	112	&	$5_17_2t_1$	&	3204	&	(1,2)	&	3209	&		\\
41	&	[$3_1t_1$,$4_18_1t_0$]	&	2414	&	(0,0)	&	2418	&	2420	&	114	&	$4_17_1t_2$	&	3223	&	(0,1)	&	3226	&		\\
42	&	$4_18_1t_0$	&	2417	&	(0,0)	&	2416	&	2415	&	115	&	[$5_27_1t_0$,$5_17_1t_2$]	&	3229	&	(2,1)	&	3235	&		\\
43	&	$3_1t_1$	&	2426	&	(0,0)	&	2427	&	2420	&	117	&	$6_28_1t_0$	&	3234	&	(1,2)	&	3232	&		\\
49	&	$7_4t_0$	&	2517	&	(1,3)	&	2517	&		&	119	&	[$5_2t_1$]	&	3247	&	(0,1)	&	3255	&		\\
50	&	$7_3t_1$	&	2530	&	(0,1)	&	2531	&	[2543]	&	120	&	$4_17_3t_0$	&	3263	&	(1,3)	&	3264	&		\\
51	&	$5_17_2t_0$	&	2558	&	(0,1)	&	2561	&	2568	&	121	&	$t_6$	&	3267	&	(0,1)	&	3295	&		\\
53	&	$5_17_1t_1$	&	2575	&	(0,0)	&	2579	&	2579	&	122	&	$4_17_2t_1$	&	3280	&	(0,2)	&	3283	&		\\
54	&	$4_1t_2$	&	2597	&	(0,0)	&	2598	&	2600	&	123	&	$8_2t_2$	&	3285	&	(1,2)	&	3290	&		\\
55	&	[$5_2t_0$,$5_1t_2$]	&	2604	&	(0,0)	&	2608	&	2608	&	124	&	$6_3t_0$	&	3291	&	(1,3)	&	3289	&	3292	\\
56	&	$4_17_2t_0$	&	2633	&	(0,1)	&	2634	&	2636	&	126	&	[$4_15_17_1t_0$]	&	3301	&	(0,1)	&	3305	&		\\
57	&	$4_17_1t_1$	&	2650	&	(0,0)	&	2653	&	2652	&	127	&	[$4_15_1t_1$]	&	3318	&	(0,1)	&	3322	&		\\
59	&	$4_15_1t_0$	&	2676	&	(0,0)	&	2678	&	2678	&	128	&	$7_28_2t_0$	&	3320	&	(2,3)	&	3320	&	3321	\\
60	&	$7_18_2t_0$	&	2690	&	(0,1)	&	2692	&	2690	&	130	&	$7_18_2t_1$	&	3331	&	(1,3)	&	3335	&		\\
61	&	$8_2t_1$	&	2703	&	(0,1)	&	2707	&	2702	&	131	&	[$8_1t_4$,$7_18_2t_1$,$8_2t_2$]	&	3337	&	(1,2)	&	3340	&		\\
63	&	$4_2t_0$	&	2746	&	(0,0)	&	2747	&	2746	&	132	&	$5_18_2t_0$	&	3365	&	(1,2)	&	3368	&		\\
64	&	$6_17_18_1t_0$	&	2764	&	(0,1)	&	2763	&	2764	&	133	&	[$8_1t_4$,$6_18_1t_2$]	&	3370	&	(1,2)	&	3374	&		\\
65	&	$6_18_1t_1$	&	2772	&	(0,1)	&	2773	&	2771	&	134	&	$4_27_1t_0$	&	3372	&	(0,1)	&	3373	&		\\
66	&	$t_5$	&	2804	&	(0,0)	&	2818	&		&	135	&	[$6_1t_4$]	&	3383	&	(1,2)	&	3386	&		\\
69	&	$3_18_1t_0$	&	2810	&	(0,0)	&	2810	&	2810	&	136	&	$6_17_28_1t_0$	&	3391	&	(2,2)	&	3389	&		\\
70	&	$8_1t_3$	&	2826	&	(0,1)	&	2830	&	2834	&	138	&	$4_2t_1$	&	3393	&	(0,3)	&	3396	&		\\
    \hline\hline\\[-0.35cm]
    \end{tabular}
}    
\end{table}

\begin{table}[H]
\singlespacing
  \caption{%
    \emph{Trans}-HCOOH. (Table~\ref{tab:trans1} continued, CVPT \cite{NeSi21}, ICPH \cite{TeMi16}.)
    \label{tab:trans3}
  } 
  \centering
\vspace{-0.4cm}  
\singlespacing
\scalebox{1.}{
  \begin{tabular}{@{}c l c@{\ }cll@{\ \ \ \ } c l c@{\ }cll@{}}
    \hline\hline\\[-0.35cm]
    \#$^\text{a}$ &	
    Assign.$^\text{b}$ & 
    $\tilde\nu$ &	
    $(\delta_{10},\delta_8)^\text{c}$	& 
    CVPT &
    ICPH  &
    \#$^\text{a}$ &	
    Assign.$^\text{b}$ & 
    $\tilde\nu$ &	
    $(\delta_{10},\delta_8)^\text{c}$	& 
    CVPT  &
    ICPH  \\
    \hline\\[-0.35cm]
140	&	$6_17_18_1t_1$	&	3400	&	(1,3)	&	3399	&		&	198	&	[$4_1t_4$]	&	3705	&	(0,1)	&	3712	&		\\
141	&	[$7_1t_5$]	&	3430	&	(1,1)	&	3448	&		&	199	&	$3_15_17_1t_0$	&	3706	&	(0,1)	&	3708	&		\\
142	&	$3_17_18_1t_0$	&	3430	&	(0,1)	&	3431	&		&	200	&	[$5_17_1t_3$,$7_1t_5$]	&	3706	&	(1,2)	&	3707	&		\\
143	&	[$6_1t_4$,$6_2t_2$]	&	3433	&	(1,2)	&	3433	&		&	201	&	$4_15_18_1t_0$	&	3711	&	(1,2)	&	3713	&		\\
144	&	[$5_16_18_1t_0$]	&	3434	&	(1,2)	&	3434	&		&	202	&	$7_18_3t_0$	&	3717	&	(2,5)	&	3719	&		\\
145	&	$3_18_1t_1$	&	3445	&	(0,1)	&	3447	&		&	204	&	$4_16_17_2t_0$	&	3730	&	(2,1)	&	3729	&		\\
146	&	$4_18_2t_0$	&	3451	&	(1,2)	&	3452	&		&	205	&	$8_3t_1$	&	3730	&	(2,4)	&	3735	&		\\
147	&	$6_27_2t_0$	&	3453	&	(2,4)	&	3450	&		&	206	&	$3_15_1t_1$	&	3730	&	(0,5)	&	3732	&		\\
149	&	[$6_27_1t_1$]	&	3458	&	(1,2)	&	3461	&		&	207	&	$4_16_17_1t_1$	&	3743	&	(1,2)	&	3745	&		\\
150	&	[$7_18_1t_3$]	&	3458	&	(1,3)	&	3461	&		&	208	&	[$7_4t_2$]	&	3745	&	(4,10)	&	3744	&		\\
154	&	[$5_16_2t_0$]	&	3490	&	(1,2)	&	3491	&		&	213	&	[$4_15_16_1t_0$]	&	3764	&	(1,2)	&	3770	&		\\
155	&	$3_16_17_1t_0$	&	3504	&	(0,1)	&	3503	&		&	214	&	[$5_3t_0$]	&	3777	&	(1,2)	&	3785	&		\\
156	&	[$3_16_1t_1$,$7_28_1t_2$]	&	3511	&	(1,2)	&	3514	&		&	215	&	$3_14_17_1t_0$	&	3781	&	(0,1)	&	3781	&		\\
157	&	[$4_16_18_1t_0$]	&	3512	&	(1,3)	&	3511	&		&	216	&	$4_28_1t_0$	&	3784	&	(0,2)	&	3785	&		\\
158	&	[$4_16_18_1t_0$]	&	3515	&	(1,2)	&	3514	&		&	217	&	[$4_17_1t_3$]	&	3792	&	(2,1)	&	3797	&		\\
159	&	[$3_16_1t_1$,$6_17_1t_3$]	&	3522	&	(1,2)	&	3514	&		&	218	&	$6_17_18_2t_0$	&	3793	&	(2,6)	&	3790	&		\\
161	&	[$5_28_1t_0$]	&	3536	&	(1,2)	&	3539	&		&	219	&	$6_18_2t_1$	&	3801	&	(9,5)	&	3801	&		\\
162	&	$3_2t_0$	&	3546	&	(0,0)	&	3545	&	3547	&	220	&	$7_5t_1$	&	3805	&	(6,2)	&	3797	&		\\
163	&	$7_48_1t_0$	&	3556	&	(4,5)	&	3550	&		&	221	&	$3_14_1t_1$	&	3805	&	(6,11)	&	3807	&		\\
164	&	[$7_2t_4$]	&	3557	&	(1,9)	&	3564	&		&	222	&	$7_6t_0$	&	3808	&	(3,14)	&	3783	&		\\
165	&	[$4_16_2t_0$]	&	3563	&	(1,3)	&		&		&	225	&	[$5_17_4t_0$]	&	3825	&	(5,10)	&	3825	&		\\
166	&	$2_17_1t_0$	&	3566	&	(1,3)	&	3568	&	3566	&	227	&	$3_18_2t_0$	&	3833	&	(1,4)	&	3835	&	3833	\\
167	&	$7_38_1t_1$	&	3567	&	(1,2)	&	3565	&		&	229	&	$4_26_1t_0$	&	3835	&	(1,8)	&	3835	&	3836	\\
168	&	$3_1t_3$	&	3568	&	(0,4)	&	3570	&		&	230	&	[$5_17_3t_1$]	&	3837	&	(2,9)	&	3841	&		\\
169	&	$4_16_2t_0$	&	3571	&	(1,2)	&	3565	&		&	234	&	$6_27_18_1t_0$	&	3858	&	(3,5)	&	3854	&		\\
170	&	$1_1t_0$	&	3576	&	(0,0)	&	3576	&	3575	&	235	&	[$8_2t_3$]	&	3859	&	(2,5)	&	3864	&		\\
171	&	$2_1t_1$	&	3578	&	(0,0)	&	3581	&	3579	&	237	&	[$6_28_1t_1$]	&	3865	&	(2,4)	&	3854	&		\\
173	&	[$6_17_2t_2$,$7_2t_4$]	&	3588	&	(1,3)	&	3592	&		&	239	&	[$5_27_1t_1$]	&	3873	&	(1,3)	&	3884	&		\\
174	&	$5_17_28_1t_0$	&	3592	&	(2,4)	&	3593	&		&	240	&	[$6_1t_5$]	&	3882	&	(1,3)	&	3890	&		\\
178	&	[$5_17_18_1t_1$]	&	3607	&	(1,4)	&	3611	&		&	247	&	$4_17_4t_0$	&	3898	&	(1,8)	&	3895	&		\\
179	&	[$5_1t_4$,$5_26_1t_0$]	&	3612	&	(1,2)	&	3615	&		&	249	&	$3_16_18_1t_0$	&	3911	&	(1,2)	&	3909	&		\\
180	&	$6_17_4t_0$	&	3622	&	(6,4)	&	3615	&		&	250	&	$4_17_3t_1$	&	3913	&	(2,5)	&	3914	&		\\
181	&	$3_17_1t_2$	&	3625	&	(0,9)	&	3626	&		&	251	&	[$7_18_2t_2$,$5_17_18_2t_0$]	&	3914	&	(3,4)	&	3916	&		\\
182	&	$6_17_3t_1$	&	3627	&	(3,9)	&	3625	&		&	254	&	[$6_3t_1$]	&	3918	&	(2,6)	&	3919	&		\\
183	&	[$4_18_1t_2$]	&	3634	&	(1,3)	&	3636	&		&	255	&	[$6_18_1t_3$]	&	3928	&	(2,4)	&	3929	&		\\
184	&	[$4_15_18_1t_0$]	&	3635	&	(1,5)	&	3638	&		&	256	&	[$4_15_17_2t_0$]	&	3930	&	(1,4)	&	3934	&		\\
185	&	$3_17_3t_0$	&	3649	&	(1,4)	&	3649	&		&	257	&	[$4_15_17_1t_1$]	&	3944	&	(1,2)	&	3951	&		\\
186	&	[$5_16_17_2t_0$]	&	3656	&	(2,5)	&	3656	&		&	258	&	$2_18_1t_0$	&	3952	&	(3,1)	&	3954	&	3952	\\
187	&	[$5_16_17_1t_1$,$6_17_1t_3$]	&	3659	&	(1,3)	&		&		&	260	&	$7_38_2t_0$	&	3955	&	(3,3)	&	3949	&		\\
189	&	$3_17_2t_1$	&	3664	&	(1,2)	&	3664	&		&	262	&	[$4_2t_2$,$4_25_1t_0$]	&	3959	&	(0,7)	&	3965	&		\\
190	&	[$4_1t_4$]	&	3672	&	(1,1)	&	3670	&		&	263	&	[$7_28_2t_1$]	&	3962	&	(3,7)	&	3963	&		\\
191	&	$4_17_28_1t_0$	&	3672	&	(1,4)	&	3672	&		&	264	&	[$4_15_2t_0$,$4_2t_2$]	&	3965	&	(1,6)	&	3977	&		\\
192	&	$t_7$	&	3679	&	(1,2)	&		&		&	265	&	[$7_28_2t_1$]	&	3969	&	(3,8)	&	3963	&		\\
194	&	[$7_3t_3$]	&	3687	&	(1,3)	&	3688	&		&	266	&	$3_16_2t_0$	&	3978	&	(1,2)	&	3975	&	3978	\\
195	&	$4_17_18_1t_1$	&	3688	&	(1,3)	&	3690	&		&	268	&	[$6_2t_3$]	&	3985	&	(2,4)	&	3984	&		\\
196	&	[$5_16_1t_2$]	&	3693	&	(1,2)	&	3698	&		&	269	&	$5_17_18_2t_0$	&	3993	&	(3,6)	&	3994	&		\\
    \hline\hline\\[-0.35cm]
    \end{tabular}
}    
  \vspace{-0.1cm}
  \begin{flushleft}
    $^\text{a}$ \#: Number of the state in the full vibrational energy list including \cis, \trans, and \emph{delocalized} states. \\
    $^\text{b}$ Excitation number for the \rc-normal modes, $1_n,2_n,3_n,4_n,5_n,6_n,7_n,8_n\ (n=0,1,\ldots)$, zero excitation is not shown. For the 9th degree of freedom, 
    the type of torsional function (Fig.~\ref{fig:cistrans}) and the excitation number is indicated, $t_n/c_n/d_n$ with $n\geq0$. `[...]' labels the largest contribution(s) from strongly mixed states. The symmetry behaviour with respect to plane reflection can be derived from the properties of the 1D basis functions according to Table~\ref{tab:planesym} by multiplication of the characters. \\
    $^\text{c}$ $(\delta_{10},\delta_8)=(\tilde\nu_{b=9}-\tilde\nu_{b=10},\tilde\nu_{b=8}-\tilde\nu_{b=9})$ is  shown for assessment of the convergence. The reported $\tilde\nu$ values and the assignment correspond to the $b=9$ basis set. 
    %
    %
    %
    %
    %
  \end{flushleft}
\end{table}

\begin{table}[H]
\singlespacing
  \caption{%
    \emph{Cis}-HCOOH (including contributions from \trans\ and \emph{delocalized} states): vibrational energies, $\tilde\nu$ in \cm\  referenced to the \trans-ZPVE, 
    computed with the \gesm\ approach in comparison with the
    internal-coordinate reaction-path Hamiltonian (ICPH) results \cite{TeMi16}.
    (The footnotes are below Table~\ref{tab:cis2}.)  
    \label{tab:cis1}
  } 
  \centering
\vspace{-0.4cm}  
\singlespacing
\scalebox{1.}{
  \begin{tabular}{@{}c l c@{\ }cl@{\ \ \ \ } c l c@{\ }cl@{}}
    \hline\hline\\[-0.35cm]
    \#$^\text{a}$ &	
    Assign.$^\text{b}$ & 
    $\tilde\nu$ &	
    $(\delta_{10},\delta_8)^\text{c}$	& 
    ICPH &
    \#$^\text{a}$ &	
    Assign.$^\text{b}$ & 
    $\tilde\nu$ &	
    $(\delta_{10},\delta_8)^\text{c}$	& 
    ICPH \\
    \hline\\[-0.35cm]
11	&	ZPV-$c_0$	&	1418	&	(0,0)	&	1412	&	245	&	[c$-$t mixed]	&	3897	&	(3,2)	&		\\
21	&	$c_1$	&	1908	&	(0,0)	&	1904	&	246	&	[c$+$t mixed]	&	3898	&	(1,4)	&		\\
27	&	$7_1c_0$	&	2076	&	(0,0)	&	2080	&	248	&	[$5_1c_3$,$6_1c_3$]	&	3902	&	(2,10)	&	[3915]	\\
38	&	$c_2$	&	2372	&	(0,0)	&	2400	&	252	&	[$c_6$$-$t mixed,$d_0$]	&	3914	&	(2,5)	&		\\
44	&	$8_1c_0$	&	2439	&	(0,1)	&	2450	&	253	&	[$c_6$$+$t mixed,$d_0$]	&	3916	&	(2,7)	&		\\
48	&	$5_1c_0$	&	2514	&	(0,0)	&	2514	&	259	&	$8_2c_1$	&	3953	&	(2,4)	&		\\
52	&	$7_1c_1$	&	2568	&	(0,1)	&	2577	&	261	&	$4_17_1c_1$	&	3956	&	(2,4)	&		\\
58	&	$6_1c_0$	&	2665	&	(0,1)	&	2667	&	267	&	[$7_3c_0$,$6_17_2c_0$]	&	3980	&	(2,8)	&		\\
62	&	$7_2c_0$	&	2733	&	(1,1)	&	2759	&	276	&	[$5_18_1c_1$,$6_18_1c_1$]	&	4027	&	(3,5)	&		\\
67	&	$4_1c_0$	&	2807	&	(0,0)	&	2806	&	277	&	[$6_1c_3$$+$$3_18_1t_2$]	&	4029	&	(3,7)	&		\\
68	&	$c_3$	&	2809	&	(1,1)	&		&	278	&	[$6_1c_3$$-$$3_18_1t_2$]	&	4029	&	(2,7)	&		\\
78	&	$8_1c_1$	&	2931	&	(1,2)	&	2937	&	281	&	[$d_2$,c,t mixed]	&	4036	&	(2,7)	&		\\
87	&	$5_1c_1$	&	3005	&	(1,1)	&	3009$^\dagger$	&	283	&	$4_16_1c_0$	&	4049	&	(1,2)	&	4061	\\
90	&	$7_1c_2$	&	3034	&	(1,1)	&		&	286	&	$7_4c_0$	&	4054	&	(2,8)	&		\\
99	&	$7_18_1c_0$	&	3098	&	(1,1)	&	3133	&	299	&	[$5_16_1c_1$,$5_2c_1$]	&	4096	&	(2,9)	&		\\
104	&	$6_1c_1$	&	3149	&	(1,3)	&	3159$^\dagger$	&	306	&	$7_18_2c_0$	&	4116	&	(2,7)	&		\\
108	&	[$5_17_1c_0$,$6_17_1c_0$]	&	3166	&	(1,2)	&	3179	&	312	&	[$5_17_1c_2$,$6_17_1c_2$]	&	4135	&	(2,3)	&		\\
113	&	$c_4$	&	3215	&	(1,1)	&		&	314	&	$7_2c_3$	&	4142	&	(4,6)	&		\\
116	&	$7_2c_1$	&	3229	&	(0,2)	&		&	321	&	$4_2c_0$	&	4166	&	(0,2)	&	4170	\\
118	&	$3_1c_0$	&	3239	&	(0,0)	&	3235	&	324	&	[$6_18_1c_1$,$5_18_1c_1$]	&	4178	&	(1,3)	&		\\
125	&	$4_1c_1$	&	3296	&	(0,1)	&	3299	&	327	&	[$5_17_18_1c_0$,$6_17_18_1c_0$]	&	4187	&	(3,5)	&		\\
129	&	[$6_17_1c_0$]	&	3321	&	(1,3)	&		&	328	&	$3_1c_2$	&	4189	&	(2,4)	&		\\
137	&	$7_3c_0$	&	3391	&	(1,4)	&		&	329	&	[c,t mixed]	&	4189	&	(1,4)	&		\\
139	&	$8_1c_2$	&	3398	&	(1,4)	&		&	332	&	$4_1c_3$	&	4192	&	(2,5)	&		\\
148	&	$8_2c_0$	&	3455	&	(1,4)	&	3486	&	349	&	[$6_2c_1$,$5_2c_1$]	&	4243	&	(5,11)	&		\\
151	&	$4_17_1c_0$	&	3465	&	(0,1)	&	3475	&	350	&	[$7_1c_5$,$8_1c_4$]	&	4247	&	(2,7)	&		\\
152	&	$5_1c_2$	&	3467	&	(1,3)	&	3638	&	352	&	[$5_16_17_1c_0$,$5_27_1c_0$]	&	4250	&	(1,7)	&		\\
153	&	$7_1c_3$	&	3473	&	(1,4)	&		&	353	&	$3_18_1c_0$	&	4254	&	(1,3)	&	4274	\\
160	&	$5_18_1c_0$	&	3531	&	(1,2)	&	3536	&	355	&	[$7_28_1c_1$$-$$4_16_1t_3$]	&	4259	&	(4,4)	&		\\
172	&	$c_5$	&	3582	&	(2,2)	&	3570$^{\ast,\dagger}$	&	356	&	[$7_28_1c_1$$+$$4_16_1t_3$]	&	4259	&	(4,9)	&		\\
175	&	$7_18_1c_1$	&	3594	&	(3,3)	&		&	357	&	[$7_1c_5$,$8_1c_4$]	&	4260	&	(3,8)	&		\\
176	&	[$5_16_1c_0$,$5_2c_0$]	&	3601	&	(1,3)	&	3595$^\dagger$	&	362	&	[$6_17_1c_2$,$7_2c_2$]	&	4273	&	(10,10)	&		\\
177	&	[$6_1c_2$,$5_1c_2$]	&	3605	&	(2,6)	&	3638	&	369	&	$2_1c_0$	&	4291	&	(1,10)	&	4292	\\
188	&	[$5_17_1c_1$,$7_18_1c_1$]	&	3661	&	(2,3)	&		&	372	&	[c,t mixed]	&	4299	&	(1,8)	&		\\
193	&	$6_18_1c_0$	&	3686	&	(2,3)	&	3715$^\dagger$	&	374	&	[$5_1c_4$,$6_1c_4$]	&	4303	&	(2,13)	&		\\
197	&	$7_2c_2$	&	3698	&	(4,4)	&		&	380	&	[$5_17_2c_1$,$6_17_2c_2$]	&	4320	&	(3,9)	&		\\
203	&	$3_1c_1$	&	3727	&	(0,1)	&	3733	&	383	&	[c,t mixed]	&	4325	&	(2,7)	&		\\
209	&	[$6_2c_0$,$5_2c_0$]	&	3753	&	(2,4)	&	3761$^{\ast,\dagger}$	&	384	&	$4_18_1c_1$	&	4325	&	(2,7)	&		\\
210	&	$4_1c_2$$-$[t mixed]	&	3757	&	(2,2)	&		&	387	&	[$3_15_1c_0$,$3_16_1c_0$]	&	4330	&	(1,4)	&	4322	\\
211	&	$4_1c_2$$+$[t mixed]	&	3757	&	(1,4)	&		&	391	&	[$6_17_18_1c_0$,$7_28_1c_0$]	&	4346	&	(4,10)	&		\\
212	&	$7_28_1c_0$	&	3758	&	(1,6)	&		&	398	&	$7_3c_2$	&	4366	&	(5,12)	&		\\
223	&	$6_17_1c_1$	&	3811	&	(5,14)	&		&	399	&	[$d_3$,c,t mixed]	&	4371	&	(5,9)	&		\\
224	&	[$5_17_2c_0$,$6_17_2c_0$]	&	3820	&	(2,15)	&		&	404	&	[$6_2c_1$,$5_16_1c_1$]	&	4378	&	(2,7)	&		\\
226	&	[c,t mixed]	&	3833	&	(2,4)	&		&	405	&	$3_17_1c_1$	&	4382	&	(2,6)	&		\\
228	&	$4_18_1c_0$	&	3833	&	(1,4)	&	3854	&	407	&	[$3_17_1c_1$,$4_15_1c_1$]	&	4388	&	(5,3)	&		\\
231	&	[$8_1c_3$,$8_1t_5$]	&	3842	&	(2,12)	&		&	416	&	[$6_27_1c_0$,$5_17_2c_0$]	&	4408	&	(5,10)	&		\\
232	&	[c$+$t mixed]	&	3850	&	(1,7)	&		&	419	&	[c$-$t mixed]	&	4420	&	(4,5)	&		\\
233	&	[c$-$t mixed]	&	3854	&	(1,4)	&		&	420	&	[c$+$t mixed]	&	4420	&	(3,9)	&		\\
236	&	[c,t mixed]	&	3860	&	(2,5)	&		&	421	&	$4_17_1c_2$	&	4421	&	(4,10)	&		\\
238	&	[c,t mixed]	&	3873	&	(1,2)	&		&	422	&	$7_38_1c_0$	&	4423	&	(5,10)	&		\\
241	&	$7_1c_4$	&	3883	&	(2,6)	&		&	424	&	$8_2c_2$	&	4427	&	(--,9)	&		\\
242	&	[$3_17_1c_0$,$4_15_1c_0$]	&	3890	&	(2,2)	&		&	437	&	$8_3c_0$	&	4467	&	(--,8)	&		\\
243	&	$7_3c_1$	&	3891	&	(3,4)	&		&	441	&	[$7_3c_1$,$6_17_2c_1$]	&	4477	&	(--,12)	&		\\
244	&	[$3_17_1c_0$,$6_2c_0$]	&	3892	&	(2,7)	&	3908	&	444	&	[$5_27_2c_0$,$6_17_3c_0$,$5_17_3c_0$]	&	4479	&	(--,15)	&	\\
    \hline\hline\\[-0.35cm]
    \end{tabular}
}    
\end{table}

\begin{table}[H]
\singlespacing
  \caption{%
    \emph{Cis}-HCOOH (including contributions from \trans\ and \emph{delocalized} states). (Table~\ref{tab:cis1} continued, ICPH \cite{TeMi16}.)
    \label{tab:cis2}
  } 
  \centering
\vspace{-0.4cm}  
\singlespacing
\scalebox{1.}{
  \begin{tabular}{@{}c l c@{\ }cl@{\ \ \ \ } c l c@{\ }cl@{}}
    \hline\hline\\[-0.35cm]
    \#$^\text{a}$ &	
    Assign.$^\text{b}$ & 
    $\tilde\nu$ &	
    $(\delta_{10},\delta_8)^\text{c}$	& 
    ICPH &
    \#$^\text{a}$ &	
    Assign.$^\text{b}$ & 
    $\tilde\nu$ &	
    $(\delta_{10},\delta_8)^\text{c}$	& 
    ICPH \\
    \hline\\[-0.35cm]
446	&	[$3_16_1c_0$]	&	4483	&	(--,16)	&	4496	&	492	&	[$7_1c_6$,c,d,t mixed]	&	4583	&	(--,8)	&		\\
447	&	[c$+$t mixed]	&	4485	&	(--,14)	&		&	496	&	[$4_1c_4$$+$$3_16_27_1t_0$]	&	4596	&	(--,7)	&		\\
448	&	[c$-$t mixed]	&	4490	&	(--,12)	&		&	497	&	[$4_1c_4$$-$$3_16_27_1t_0$]	&	4597	&	(--,7)	&		\\
452	&	$4_17_18_1c_0$	&	4493	&	(--,17)	&		&	498	&	$4_1c_4$	&	4598	&	(--,7)	&		\\
453	&	[$5_18_1c_2$,$6_18_1c_2$]	&	4495	&	(--,17)	&		&	508	&	$4_17_2c_1$	&	4618	&	(--,9)	&		\\
454	&	[c,t mixed]	&	4496	&	(--,19)	&		&	509	&	[$5_28_1c_0$,$5_16_18_1c_0$]	&	4619	&	(--,8)	&		\\
456	&	[c,t mixed]	&	4506	&	(--,15)	&		&	510	&	$7_18_2c_1$	&	4620	&	(--,15)	&		\\
459	&	[$7_18_1c_3$$+$$5_1t_6$]	&	4514	&	(--,12)	&		&	511	&	$3_14_1c_0$	&	4622	&	(--,15)	&	4626	\\
460	&	[$7_18_1c_3$$-$$5_1t_6$]	&	4518	&	(--,10)	&		&	512	&	$3_1c_3$	&	4622	&	(--,14)	&		\\
465	&	[c,t mixed]	&	4529	&	(--,8)	&		&	515	&	[$8_1c_5$$-$$4_1t_6$]	&	4629	&	(--,18)	&		\\
467	&	[$4_16_1c_1$]	&	4534	&	(--,8)	&		&	517	&	[$8_1c_5$$+$$4_1t_6$]	&	4637	&	(--,12)	&		\\
470	&	[$3_17_2c_0$]	&	4543	&	(--,11)	&		&	520	&	[$6_18_1c_2$]	&	4643	&	(--,10)	&		\\
472	&	$5_18_2c_0$	&	4546	&	(--,10)	&		&	524	&	[$7_4c_0$,$6_17_3c_0$,$5_16_17_2c_0$]	&	4646	&	(--,11)	&		\\
476	&	[$3_17_2c_0$$-$$6_37_2t_0$]	&	4550	&	(--,15)	&		&	527	&	$4_2c_1$	&	4654	&	(--,7)	&	4672	\\
477	&	[$3_17_2c_0$$+$$6_37_2t_0$]	&	4550	&	(--,16)	&		&	535	&	[$5_1c_5$]	&	4671	&	(--,7)	&		\\
479	&	[$7_2c_4$]	&	4553	&	(--,17)	&		&	542	&	[$5_3c_0$,$5_26_1c_0$]	&	4684	&	(--,17)	&		\\
480	&	[$7_2c_4$]	&	4553	&	(--,16)	&		&	544	&	[$5_17_18_1c_1$,$6_17_18_1c_2$]	&	4689	&	(--,15)	&		\\
481	&	[$5_16_1c_2$]	&	4557	&	(--,16)	&		&	547	&	[c,t mixed]	&	4697	&	(--,14)	&		\\
484	&	$7_4c_1$	&	4564	&	(--,16)	&		&	549	&	[c,t mixed]	&	4702	&	(--,12)	&		\\
485	&	[$5_167_1c_3$,$6_17_1c_3$]	&	4565	&	(--,16)	&		&	550	&	[c,t mixed]	&	4702	&	(--,12)	&		\\
\hline\hline\\[-0.35cm]
    \end{tabular}
}    
  \vspace{-0.1cm}
  \begin{flushleft}
    $^\text{a,b,c}$ see footnotes to Fig.~\ref{tab:trans3}. \\
    $^\dagger$: Revised assignment based on the Supplementary Material of Ref.~\cite{TeMi16}. \\
    $^\ast$: Tentative comparison. 
  \end{flushleft}
\end{table}

\begin{table}[H]
  \centering
  \caption{%
    \emph{Delocalized}-HCOOH: selected vibrationally excited states 
    computed with the \gesm\ approach ($b=9$) with significant contribution from delocalized torsional basis functions, $d_0,d_1,d_2,$ and $d_3$. The vibrational energy, $\tilde\nu$ in \cm, is referenced to the \trans-ZPVE.
    \label{tab:deloc}  
  }
  \begin{tabular}{@{}c@{\ \ \ }l@{}l@{}l@{}l @{\ \ \ }c@{\ \ }c@{}}
  \hline\hline\\[-0.40cm]
  \#$^\text{a}$ & 
  \multicolumn{4}{c}{Dominant basis-state contributions$^\text{b}$} & 
  $\tilde\nu$ & 
  $(\delta_{10},\delta_8)^\text{c}$ \\
  \hline
252	&	$-0.56$ $6_37_1t_0$	    &	$+0.32$ $6_47_1t_0$ 	&	$\ldots+0.20$ $c_6$ 	&	$+0.15$ $d_0$	    &	3914	& (2,5) \\
253	&	$+0.44$ $c_6$	        &	$+0.37$ $d_0$	        &  	$-0.33$ $7_1t_6$	    &	                	&	3916	& (2,7) \\
281	&	$-0.36$ $d_1$	        &	$+0.32$ $5_1t_5$	    &		                    &		                &	4036	& (2,7) \\
284	&	$+0.34$ $6_27_1t_2$	    &	$+0.32$ $d_0$            &	$-0.30$ $6_17_1t_4$	    &		                &	4053	& (5,4) \\
287	&	$+0.41$ $d_0$	        &	$+0.33$ $7_1t_6$	    &	$+0.31$ $5_16_17_18_1t_0$&		                &	4060	& (2,5) \\
303	&	$+0.41$ $d_1$	        &	$+0.30$ $3_15_18_1t_0$	&	$+0.27$ $5_26_1t_1$	    &	$-0.25$ $6_1t_3$	&	4112	& (2,4) \\
309	&	$+0.47$ $d_1$	        &	$-0.27$ $5_16_2t_1$	    &	$-0.27$ $5_26_1t_1$	    &		                &	4125	& (2,4) \\
376	&	$-0.33$ $5_3t_1$	    &	$+0.30$ $d_2$	        &	$+0.30$ $5_16_1t_3$	    &		                &	4311	& (2,10) \\
389	&	$+0.42$ $4_15_17_18_1t_0$&	$+0.25$ $4_17_18_1t_2$	&	$+0.21$ $d_2$	        &		                &	4340	& (2,6) \\
399	&	$+0.55$ $d_3$	        &	$+0.22$ $3_14_1t_2$	    &	$+0.21$ $6_1t_6$	    &		                &	4371	& (4,9) \\
  \hline\hline\\[-0.35cm]
  \end{tabular}  
  \vspace{-0.1cm}
  \begin{flushleft}
    $^\text{a,b,c}$ see footnotes to Fig.~\ref{tab:trans3}. 
  \end{flushleft}
\end{table}

%
%
\clearpage
\section{Summary, conclusion, and outlook}
\noindent
Variational vibrational excitation energies have been reported for the formic acid molecule up to ca.~4700~\cm, which is slightly beyond the top of the \cis-\trans\ isomerization barrier, using system-adapted curvilinear coordinates in the \gesm\ approach developed in the present work and a high-level \emph{ab initio} potential energy surface (PES) taken from Ref.~\cite{TeMi16}.

The results confirm (within 1--5~\cm) up to 2500~\cm, and improve (by 5--10~\cm) between 2500 and 4000~\cm\ the 6th-order canonical van Vleck perturbation theory (CVPT) energies \cite{NeSi21} that were obtained from a computation localized on the \trans\ PES well.
Both the \cis\ and \trans\ energies computed with the \gesm\ approach are in an overall good agreement, but improve (by 10--40~\cm) upon the internal-coordinate path Hamiltonian (ICPH) results that similarly to the present work account for both the \cis\ and \trans\ wells of the PES.
There exists another potential energy surface and multi-configuration time-dependent Hartree (MCTDH) computations have been reported using that PES \cite{RiCa18,AeCaRiBr20}. Direct comparison with those results have not been reported in this work, because our current focus was on the development of a computational procedure that can be used to provide benchmark quality vibrational energies up to and possibly beyond the isomerization barrier of the formic acid molecule.
We think that we have almost achieved this goal, further necessary work with larger basis sets and an improved basis pruning condition is in progress and results will be reported in future work.

Already in the present paper, \cis-\trans\ entangled states, corresponding `tunneling splittings',
and the (\cis\ or \trans) localized to \emph{delocalized} transition taking place near the top barrier were shortly discussed. Benchmark quality computed data (limited by the quality of the PES) on these interesting features will become available soon with the outlined theoretical, computational progress. We are not aware of detailed experimental data of these phenomena in HCOOH, and we look forward to developments from the experimental side.

\section*{Acknowledgment}
\noindent We thank the financial support of the Swiss National Science Foundation 
(PROMYS Grant, No.~IZ11Z0\_166525).
The authors are indebted to Tucker Carrington, Attila Császár, and their co-workers for joint work and discussions over the past decade that had resulted in ideas and developments necessary to accomplish the present research. 
We also thank Martin Suhm and Arman Nejad who made us interested in working on this system.

%
%
%

\clearpage

\section*{Supplementary Material}

\begin{table}[h!]
\caption{%
Curvilinear normal coordinate ($\xi_i$) parameters
for the \textcolor{red}{\trans} formic acid
and the $\tilde{\nu}^{\HO}$ harmonic frequencies in \cm.
$\xi_i= \xi^{\text{(eq)}}_i + \sum_{j=1}^8 \mathcal{Q}_j \mathcal{L}_{i,j}$.
The units correspond to bohr for the distances and radian for the angles.\\
\label{tab:transnormcoord}
}
\scalebox{0.65}{%
\rotatebox{0}{%
\begin{tabular}{@{}c@{\ }|rrrrrrrrrr@{}}
\hline \\[-0.7cm] \hline \\[-0.2cm]
 &        &      &      $\mathcal{Q}_{1}$     &        $\mathcal{Q}_{2}$      &       $\mathcal{Q}_{3}$     &         $\mathcal{Q}_{4}$    &        $\mathcal{Q}_{5}$      &       $\mathcal{Q}_{6}$   &       $\mathcal{Q}_{7}$ &       $\mathcal{Q}_{8}$ \\[0.2cm]
\hline \\[-0.2cm] 
$\tilde{\nu}^{\HO}$    &  &  & 3765.35 &  3089.16      &    1816.23    &     1412.08     &      1323.08     &    1140.45  &   1102.9  &    631.798     \\[0.2cm]
\hline \\[-0.2cm] 
  &    \multicolumn{1}{c}{$\xi^{\text{(eq)}}$}        & $\mathcal{L}_{i,j}$: &   &  &  &   &    &    \\[0.2cm]
\hline \\[-0.2cm] 
$r_1$      &      2.53601         & &   $-$0.00160774 &  $-$0.00280096 & $-$0.02092050 &     0.00377681 &   0.02909350 &   $-$0.05012370 &         ~~0.00000000 &   $-$0.01130860  \\[0.2cm]
$r_2$      &       2.26578        & &    0.00010330 &    $-$0.00563876 &  0.05024730 &   0.01044820 &  $-$0.00463876 & $-$0.00645253 &         0.00000000 &   $-$0.00125626  \\[0.2cm]
$r_3$      &      2.06736        & &   $-$0.00179791 &   0.10827700 &    0.00275970 &   0.00264300 &   0.00228843 &    0.00062604 &   0.00000000 &   $-$0.00061152  \\[0.2cm]
$r_4$      &      1.82737         & &    0.09711080 &     0.00200363 &   $-$0.00081998 &    0.00186474 &  $-$0.00447978 & $-$0.00239879 &         0.00000000 &    0.00087992    \\[0.2cm]
$\theta_1$ &       2.17992        & &    0.00180103 &     0.00647238 &   $-$0.00956847 &    $-$0.00088388 &       $-$0.02380230 &  0.01197050 &   0.00000000 &   $-$0.04093320  \\[0.2cm]
$\theta_2$ &       1.91986        & &   $-$0.00193900 &  $-$0.00262074 &  0.03055210 &  $-$0.06940420 &    $-$0.01605430 & $-$0.01774360 &         0.00000000 &    0.02458280    \\[0.2cm]
$\theta_3$ &       1.86227        & &   $-$0.00028928 &  $-$0.00389619 &  0.02221480 &  $-$0.02777300 &     0.06806850 &    0.05177930 &   0.00000000 &   $-$0.02681930  \\[0.2cm]
$\tau_1$   &      0.00000           & &    0.00000000 &     0.00000000 &    0.00000000 &   0.00000000 &   0.00000000 &    0.00000000 &   0.12015200 &    0.00000000    \\[0.4cm]

\hline \\[-0.7cm] \hline \\[-0.2cm] 
\end{tabular} 
}}
\end{table}     

\begin{table}[h!]
\caption{%
Curvilinear normal coordinate ($\xi_i$) parameters 
for the \textcolor{red}{\cis} formic acid
and the $\tilde{\nu}^{\HO}$ harmonic frequencies in \cm.
$\xi_i= \xi^{\text{(eq)}}_i + \sum_{j=1}^8 \mathcal{Q}_j \mathcal{L}_{i,j}$.
The units correspond to bohr for the distances and radian for the angles.\\
\label{tab:cisnormcoord}
}
\scalebox{0.65}{%
\rotatebox{0}{%
\begin{tabular}{@{}c@{\ }|rrrrrrrrrr@{}}
\hline \\[-0.7cm] \hline \\[-0.2cm]
 &        &      &      $\mathcal{Q}_{1}$     &        $\mathcal{Q}_{2}$      &       $\mathcal{Q}_{3}$     &         $\mathcal{Q}_{4}$    &        $\mathcal{Q}_{5}$      &       $\mathcal{Q}_{6}$   &       $\mathcal{Q}_{7}$ &       $\mathcal{Q}_{8}$ \\[0.2cm]
\hline \\[-0.2cm] 
$\tilde{\nu}^{\HO}$    &  &  & 3826.32 &  3006.92      &    1861.52    &     1428.97     &      1300.17    &   1125.37  &   1040.96  &     664.495     \\[0.2cm]
\hline \\[-0.2cm] 
  &    \multicolumn{1}{c}{$\xi^{\text{(eq)}}$}        & $\mathcal{L}_{i,j}$: &   &  &  &   &    &    \\[0.2cm]
\hline \\[-0.2cm]
$r_1$      &      2.54924         & &   $-$0.00208335  &  $-$0.00284458 &  $-$0.01799320  & 0.00497105 &  0.03407310 & $-$0.04825930  &  ~~0.00000000 &  $-$0.01260780  \\[0.2cm]
$r_2$      &       2.25255        & &   0.00036668  &  $-$0.00571702 & 0.04977190  & 0.00837307 & $-$0.00669357 & $-$0.00607253  &  0.00000000 &  $-$0.00269151  \\[0.2cm]
$r_3$      &      2.07870        & &   $-$0.00213001  & 0.10972700 & 0.00329455 & 0.00295397 &  0.00078314 &  0.00230952  &  0.00000000 &  $-$0.00087318  \\[0.2cm]
$r_4$      &      1.81792         & &  0.09636670  &  0.00240434 & $-$0.00130866 &  0.00061663 &  $-$0.00231159 &  $-$0.00001521  &  0.00000000 &  0.00012556  \\[0.2cm]
$\theta_1$ &       2.13454       & &   0.00179629  & 0.00633024 &  $-$0.00979878 &  $-$0.00421724 & $-$0.01876390 &  0.01536720  &  0.00000000 &  $-$0.04226280  \\[0.2cm]
$\theta_2$ &      1.98444       & &  $-$0.00175648  &  $-$0.00257378 &  0.02977090  &  $-$0.07143720 & $-$0.00623438 & $-$0.01340060  &  0.00000000 & 0.02285910  \\[0.2cm]
$\theta_3$ &       1.90590        & &   $-$0.00118818  & $-$0.00387435 &  0.02117820 & $-$0.01419170 & 0.08259160 & 0.03606030  &  0.00000000 & $-$0.02161040  \\[0.2cm]
$\tau_1$   &      0.00000           & &     0.00000000  &  0.00000000 &  0.00000000 &  0.00000000 &  0.00000000 &  0.00000000  &  0.12498000 &  0.00000000  \\[0.4cm]
\hline \\[-0.7cm] \hline \\[-0.2cm] 
\end{tabular} 
}}
\end{table} 

\end{document}